\documentclass[a4paper, 12pt]{article}
\usepackage{amstext}
\usepackage{amsmath}
\usepackage{amsthm}
\usepackage{mathtools}
\usepackage{physics}
\usepackage{mathrsfs}
\usepackage{array}
\usepackage{titling}
\usepackage{enumitem}
\usepackage[dvipsnames]{xcolor}
\usepackage[calc, en-US]{datetime2}

\usepackage[labelsep=period]{caption}
\usepackage{subcaption}

\usepackage[page]{appendix}

\usepackage{graphicx}
\graphicspath{{graphs/}}

\usepackage[margin=1.25in]{geometry}
\usepackage[onehalfspacing]{setspace}

\usepackage[scaled=1,helvratio=1]{newtxtext}
\usepackage[varg,vvarbb]{newtxmath}
\usepackage[varqu,varl]{inconsolata}
\usepackage[type1,scaled=1]{cabin}
\DeclareMathAlphabet{\mathcal}{OMS}{cmsy}{m}{n}

\usepackage{anyfontsize}
\usepackage{bm}

\DeclareMathOperator{\ind}{\mathbf{1}}
\DeclareMathOperator*{\argmax}{arg\,max}

\newcommand{\R}{\mathbb{R}}
\newcommand{\ehat}{\hat{\mathbf{e}}}

\DeclareMathOperator{\interior}{\mathrm{int}}

\theoremstyle{plain}
\newtheorem{theorem}{Theorem}
\newtheorem{lemma}{Lemma}
\newtheorem{proposition}{Proposition}

\newtheorem{assumption}{Assumption}

\theoremstyle{definition}
\newtheorem{definition}{Definition}

\newtheorem{example}{Example}

\theoremstyle{remark}

\usepackage[
backend=biber,
sorting=nyt,
citestyle=authoryear,
bibstyle=authortitle,
date=year,
doi=false,
isbn=false,
url=false,
eprint=false
]{biblatex}
\usepackage{hyperref}
\hypersetup{
    colorlinks,
    linkcolor = Blue,
    citecolor = Blue,
    urlcolor = Black
}

\pretitle{\begin{center}\LARGE\bfseries}
\posttitle{\par\end{center}\vskip 1.5em

\par
}

\preauthor{\begin{center}
    \large \vskip 0.5em
    }
\postauthor{\par\end{center}}

\predate{\begin{center}\small\normalfont\itshape}
\postdate{\par\end{center}}

\title{Optimal Design of Climate Disclosure Policies: Transparency versus Externality}
\author{
    Shangen Li\thanks{
        \protect Zurich Center for Market Design,
University of Zurich
(e-mail address: \url{work@shangen.li}).
        \protect I would like to thank
Marek Pycia,
Armin Schmutzler,
Haoyuan Zeng, 
and seminar participants at
Swiss Theory Day 2023
and the 15th Workshop on Accounting \& Economics
for helpful comments.

\includegraphics[trim={2cm 5cm 2cm 5cm}, clip, height=\baselineskip]{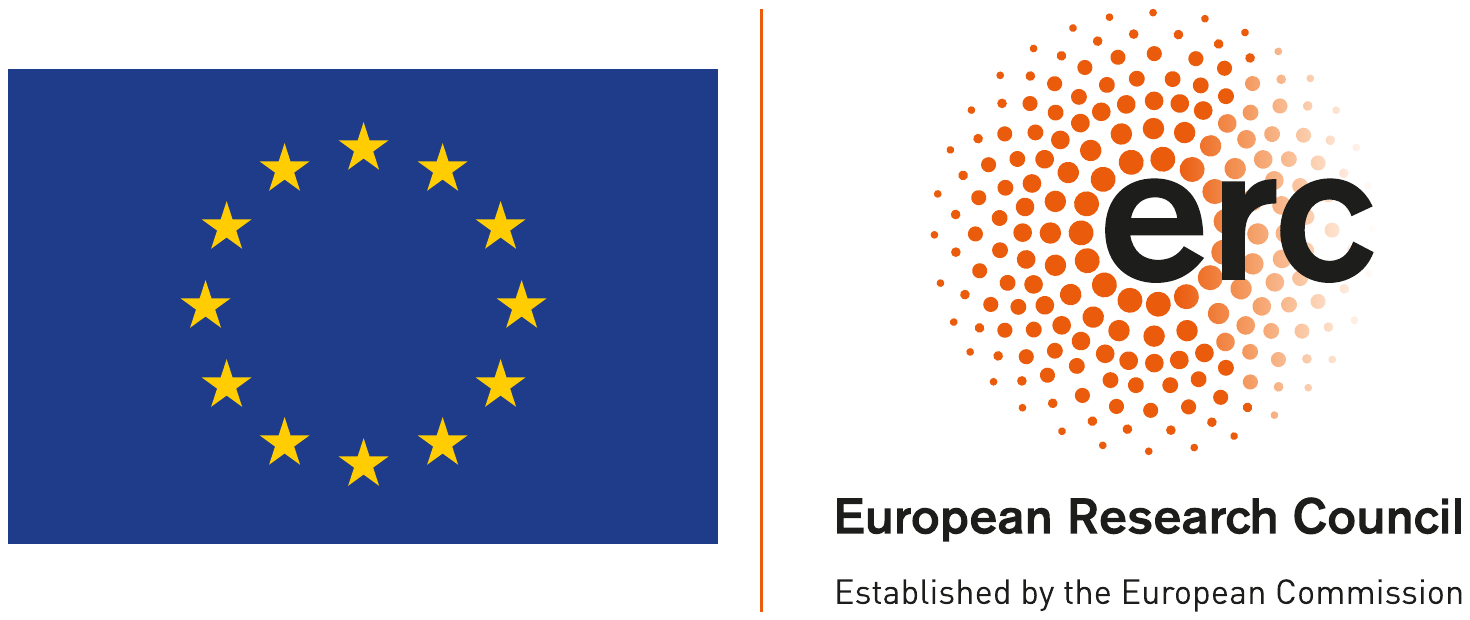} 
This project has received funding from the European Research Council (ERC) under the European Union's Horizon 2020 research and innovation program (Grant agreement No.\ 866376).
    }
}

\date{\DTMdate{2024-8-20}}

\begin{document}

\maketitle

\begin{abstract}
    Does a more transparent climate disclosure policy induce lower emissions?
    This paper examines the welfare implications of transparency in climate disclosure regulation.
    Increased disclosure transparency
    could result in a larger equilibrium externality,
    but never leaves the firm worse off.
    Consequently,
    mandating full disclosure is no different from maximizing the firm's private benefit
    while disregarding the ensuing externality.
    Transparency beyond binary disclosure is necessary
    only when 
    the firm holds private information about
    its incentives for emission reduction.
    I provide conditions under which 
    focusing on
    threshold disclosure policies
    entails no loss of generality.
\end{abstract}

{\bfseries Keywords:} Sustainable finance, Carbon disclosure, Information design.

{\bfseries JEL classification:} G18; M48; Q58.

\section{Introduction}

In recent years,
there has been an unprecedented surge in public awareness of climate impacts, 
with the financial sector being no exception.
Investors have been showing a growing interest in scrutinizing the environmental impacts of the firms they consider for investment,
calling for greater transparency in the disclosure of corporate carbon footprints.
In response to this demand, the regulation of corporate climate disclosure has taken center stage on the agenda of the policymakers. 
From a welfare perspective,
disclosing firms' climate-related activities,
such as their greenhouse gas emissions,
serves to hold these firms accountable for their actions,
and thereby facilitates the internalization of their externalities.\footnote{
    The idea that better monitoring mitigates moral hazard
    can be traced back at least to \textcite{Holmstrom:1979}.
}
Nevertheless,
somewhat paradoxically,
more disclosure does not always lead to a reduction in externalities,
as demonstrated in the following example.\footnote{
    \textcite{Dewatripont-Jewitt-Tirole:1999} 
    provide an analogous example within the context of grading scale design in schools.
    They show that a pass/fail grading scheme can induce greater effort from students than a system with fully disclosed grades.
}

Consider a firm raising capital for a project
that involves a choice among three levels of emission: \(\{\texttt{LOW}, \texttt{MID}, \texttt{HIGH}\}\).
Perhaps due to the costs associated with emission reduction,
let us suppose the revenue of the project increases with the chosen emissions,
given by \(R(\texttt{LOW})=3\), \(R(\texttt{MID})=5\) and \(R(\texttt{HIGH}) = 6\).
To finance this project,
the firm seeks investment in the financial market.
As a consequence of the investors' preference for green assets,
the cost of capital increases with the firm's emission level \emph{perceived} by the market,\footnote{
    \textcite{Chava:2014} finds that firms with environmental concerns face a higher cost of capital for both equity and debt financing.
}
and is given by \(C(\texttt{LOW}) = 1\), \(C(\texttt{MID}) = 2\) and \(C(\texttt{HIGH}) = 5\).
Furthermore,
assume that
the firm's emissions can only be verified based on the disclosure policy prescribed by the regulator.\footnote{
    Motivated by the well-known challenges of verifying carbon footprints in practice, 
    this paper posits that effective verification can only be achieved through regulatory disclosure.
    Firms are expected to comply with these regulations
    and report their emissions truthfully,
    given the severe penalties imposed for any fraudulent disclosure.
}
With such a disclosure policy in place,
the firm then chooses the level of emission to maximize its profit.

The benefit of disclosure is straightforward.
In the absence of such a disclosure policy,
no avenue is available for the firm to prove its emission level to the investors.
As emission reduction is costly,
the investors would never believe that a \texttt{LOW} or \texttt{MID} level of emission has been chosen.
As a consequence,
by choosing the emission level \texttt{HIGH},
the firm earns a profit of \(R(\texttt{HIGH}) - C(\texttt{HIGH}) = 1\).
By contrast, under a disclosure policy that fully discloses emissions,
the firm would choose the emission level \texttt{MID},
obtaining a profit of \(R(\texttt{MID}) - C(\texttt{MID}) = 3\).
Hence,
a disclosure mandate
not only helps internalize the externality
but also raises the firm's profit,
thereby leading to a more efficient outcome.

Can there be a disclosure policy that incentivizes the firm to lower its emissions even further?
Consider the policy that
discloses the level \texttt{LOW} as before,
but pools the levels \texttt{MID} and \texttt{HIGH} together.
As the market cannot distinguish \texttt{MID} from \texttt{HIGH}
under such a disclosure policy,
the firm would never choose \texttt{MID},
as it incurs a cost of capital the same as \texttt{HIGH},
yet delivers a lower revenue.
Given that \texttt{HIGH} emission yields a profit of 1,
the firm would then choose the emission level \texttt{LOW},
which yields a profit of \(R(\texttt{LOW}) - C(\texttt{LOW}) = 2\).

This example demonstrates a trade-off between transparency and externality in the design of climate disclosure policies.
Transparency sometimes reduces emissions,
but not always:
full disclosure induces an emission level lower than no disclosure,
but not as low as the less transparent policy that pools \texttt{MID} and \texttt{HIGH}.

This paper studies the design of climate disclosure policies,
with a particular focus on the role of disclosure transparency.
The model in Section~\ref{sec:model} formalizes the example above and generalizes it in two aspects as follows.
First, the emission space is taken to be a closed interval in the real line instead of a set of discrete values as above.
Second and more importantly,
in addition to moral hazard,
the model also features adverse selection.
The revenue of the project depends not only on the chosen emissions,
but also on the firm's marginal incentives for emission reduction,
which is assumed to be the firm's private information.

In a benchmark setting
where the regulator and the firm are symmetrically informed,
any individually rational emission level
can be implemented by
a binary disclosure policy
that discloses
whether or not the firm's emission exceeds a critical value.
Because
the equilibrium emission level can be anticipated under each given policy,
further transparency is deemed unnecessary under symmetric information.

Under asymmetric information,
disclosure policy also serves as a screening device
apart from its role of monitoring,
with more transparent disclosure being able to screen the firm's type more excessively.
Therefore,
policies more transparent than binary disclosure
may lead to Pareto improvements,
in marked contrast to the benchmark scenario.
This observation reveals that in the context of climate disclosure,
transparency is desired mainly because of its advantage in tackling adverse selection,
rather than,
as suggested by conventional wisdom,
moral hazard.

My analysis then proceeds to study the effect of transparency on welfare.
In contrast to the transparency versus externality trade-off illustrated earlier,
it turns out that more transparent disclosure always makes the \emph{firm} weakly better off.
The intuition behind this seemingly counterintuitive statement is that
more transparent disclosure
offers the firm a richer set of emission levels that can be sustained in equilibrium,
and thereby enlarges the feasible set of its optimization problem.
This result immediately implies that
full disclosure is Pareto efficient,
yet
induces the \emph{highest} expected emission among all efficient policies.
Put differently,
pursuing maximal transparency
is no different from maximizing the firm's profit
without concerning the ensuing externalities.

These findings naturally lead to the question of how to design disclosure policies under asymmetric information.
I characterize the Pareto frontier
of the set of all feasible emission-profit pairs
for a relatively broad class of profit functions and type distributions.
Specifically,
when the type distribution is log-concave,
any Pareto-efficient disclosure policy
can be implemented using a threshold disclosure policy:
emissions are fully disclosed if they fall weakly below a certain threshold
and are otherwise pooled together.
Furthermore, I establish conditions under which
the full-disclosure policy Pareto dominates all other policies.
Specifically,
when
the disciplining force of the market is relatively weak,
the Pareto frontier reduces to a single point,
thereby leaving full disclosure as the only remaining efficient policy.

There is a large body of literature on financial disclosure in general,
and rapidly emerging empirical studies on climate disclosure due to the nascent public interest in sustainable finance.
Recent empirical studies
(e.g., \textcite{ChenHungWang:2018};
\textcite{JouvenotKrueger:2021};
\textcite{DownarErnstbergerReichelsteinSchwenenZaklan:2021};
\textcite{YangMullerLiang:2021};
\textcite{Bolton-Kacperczyk:2021};
\textcite{Tomar:2023})
have examined the real effect of climate-related disclosure.
However, theoretical research in climate disclosure remains notably scarce.
This paper provides one of the first theoretical analyses of climate disclosure transparency and its environmental consequences.
Recently, \textcite{Xue:2023} also contributed to this line of work
by exploring a closely related design problem.
He analyzes a noisy rational expectation equilibrium model
in which disclosure policy controls the precision
of the observed emission,
which in turn
disciplines the firm
through its stock price.
He characterizes the optimal disclosure precision
that balances the firm's private benefit and the social cost of emission,
nonetheless from the standpoint of a risk-averse representative investor.
Notably,
one of his findings---less disclosure is deemed desirable when faced with stronger environmental concerns---aligns with the transparency versus externality trade-off highlighted here.
Beyond the single-firm scenario,
he also considers a large economy populated by firms with free-riding incentives.
Compared to his work,
my analysis elucidates the non-monotonic relationship between disclosure transparency and externality,
and also examines its welfare implications under adverse selection.

This paper also contributes to an emerging strand of literature on information design with moral hazard.
\textcite{Boleslavsky-Kim:2021} study a three-player Bayesian persuasion game
in which a sender designs a signal to influence a receiver's belief about an agent's hidden effort.
They characterize the optimal signal structure in a general environment,
and provide more concrete results when the state space or action space is further restricted.
\textcite{Rodina:2018} studies a similar career concerns model
and focuses on the comparative statics of incentive provision with respect to the informativeness of the signal.
Neither of these two papers considers adverse selection,
yet their settings still substantially differ from my benchmark scenario
because of the rich information structure considered in their papers.

In a closely related paper,
\textcite{Rodina-Farragut:2018} study the design of
optimal grading scheme in schools under various information environments.
The designer in their model
aims to maximize the agents's effort
without concerning their welfare.
Thus, their characterization of optimal deterministic grading schemes
essentially corresponds to the emission-minimizing disclosure policy in my model,
albeit the approach they take differs significantly from mine.

From a modeling perspective,
this paper employs the Lagrangian technique originally developed by \textcite{Amador-Bagwell:2013} for analyzing delegation problems.
\textcite{Amador-Bagwell:2022} extends this method to incorporate an ex~post participation constraint.
Building upon their method,
I derive interpretable sufficient conditions
under which the Pareto frontier is characterized by threshold policies.
The inherent connection between disclosure and delegation
leveraged in this paper
has been previously noted by
\textcite{Kolotilin-Zapechelnyuk:2019}
and
\textcite{Zapechelnyuk:2020}.

\section{Model}
\label{sec:model}

There is a firm, a regulator, and a competitive financial market.
The firm seeks investment for a project,
which yields a deterministic revenue
and involves choosing an emission level
\(e\in E\coloneqq[0, \bar{e}]\subset\mathbb{R}_+\).
At the outset of the game,
the regulator commits to a disclosure policy.
A disclosure policy is a partition of \(E\),
and can be represented by a function \(d\colon E\to E\)
that maps the firm's actual emission level \(e\)
to a disclosed level of emission \(d(e)\).

I do not model individual investors in the financial market explicitly.
Instead, I assume that after observing the disclosed emission level \(d(e)\),
the market forms a belief \(\tilde{e}\) about the firm's emission level \(e\).
The market belief \(\tilde{e}\) determines the cost of capital,
which in turn, together with the actual emission \(e\),
determines the firm's profit.
The firm's profit function \(\tilde{\pi}(\theta,e,\tilde{e})\)
is commonly known to all players,
where \(\theta\)
shapes its marginal incentives for emission reduction.
The firm's type \(\theta\)
is its private information,
and has a commonly known density \(f = F'\)
continuous on its support \(\Theta\coloneqq [\underline{\theta}, \bar{\theta}]\subset\mathbb{R}\).
For each type \(\theta\),
I assume \(\tilde{\pi}\) is strictly increasing in \(e\) and strictly decreasing in \(\tilde{e}\),
reflecting the fact that reducing emissions is costly
and higher perceived emission impedes fundraising.\footnote{
    The perceived emission levels could enter the firm's objective function 
    for various reasons,
    including corporate social responsibility, shareholders' preferences,
    or even managerial career concerns.
    I will adhere to the cost-of-capital interpretation for the moment,
    and revisit these alternative interpretations in Section~\ref{sec:optimal-policy}.
}

\paragraph*{Timing:}
The events occur in the following order.
\begin{enumerate}
    \item The regulator commits to a disclosure policy \(d\);
    \item The firm obtains capital from the financial market
    at a cost conditional on the perceived emission level \(\tilde{e}\);
    \item The firm privately chooses an emission level \(e\in E\);
    \item The market observes the disclosed emission level \(d(e)\), and forms a belief \(\tilde{e}\) about the firm's emission level;
    \item The firm earns a profit of \(\tilde{\pi}(\theta, e, \tilde{e})\).
\end{enumerate}

Let \(\pi(\theta, e) \coloneqq \tilde{\pi}(\theta, e, e)\),
which can be interpreted as the profit of the firm
when emissions are verifiable.
In addition to the monotonicity assumption on \(\tilde{\pi}\),
I further make the following regularity assumptions.
\begin{assumption}
    \label{assump:regularity}

    \begin{enumerate}
        \item \(\tilde{\pi}\in C^0(\Theta\times E\times E)\cap C^2(\interior(\Theta\times E\times E))\);
        \item \(\pi(\theta, \,\cdot\,)\) is strictly concave on \(E\) for each \(\theta\in\Theta\);
        \item \(\pi(\theta, 0) < \pi(\theta, \bar{e})\) for all \(\theta\in\Theta\).

    \end{enumerate}

\end{assumption}

\begin{definition}[Equilibrium]
    Given a disclosure policy \(d\),
    an \emph{equilibrium} \((\mathbf{e},\tilde{\mathbf{e}})\) consists of a pair of mappings 
    \(\mathbf{e}:\Theta\to E\) and \(\tilde{\mathbf{e}}:E\to E\),
    such that
    \begin{itemize}
        \item 
        anticipating the market's belief \(\tilde{\mathbf{e}}(d(e))\),
        the firm with type \(\theta\)
        maximizes its profit \(\tilde{\pi}(\theta, e, \tilde{\mathbf{e}}(d(e)))\)
        by choosing the emission level \(e=\mathbf{e}(\theta)\);
        \item 
        after observing the disclosed emission level \(d(e)\),
        the market forms its belief \(\tilde{\mathbf{e}}(d(e))\)
        according to Bayes' rule whenever possible,
        with forward induction refinement off the equilibrium path;
        \item 
        and the belief is correct: \(\mathbf{e}(\theta) = \tilde{\mathbf{e}}(d(\mathbf{e}(\theta)))\).
    \end{itemize}
\end{definition}
The forward induction refinement implies that
in any equilibrium,
upon observing an off-path disclosed emission \(d(e')\),
the investors conjecture that the firm has chosen the highest emission consistent with the disclosed information,
and form their beliefs accordingly.
Formally,
an emission level \(e\in E\) is said to be \emph{belief-compatible}
if
\(e \geq e'\) for all \(e'\in E\) such that \(d(e') = d(e)\).
Since an equilibrium emission level must be belief-compatible,
the firm's problem
\(\max_{e\in E}\{\tilde{\pi}(\theta, e, \tilde{\mathbf{e}}(d(e)))\}\) 
can be simplified to \(\max_{e\in \tilde{E}_d}\pi(\theta, e)\),
where \(\tilde{E}_d\) denotes the set of all belief-compatible emission levels.
In what follows,
let \(\mathcal{D}\) denote the set of disclosure policies 
under which an equilibrium exists.\footnote{
    It is not difficult to construct a disclosure policy in which
    an equilibrium does not exist (e.g., a disclosure policy such that
    for some type \(\theta\),
    \(\sup_{e\in E}\{\tilde{\pi}(\theta, e, \tilde{\mathbf{e}}(d(e)))\}\) cannot be attained by any \(e\in E\)).
    One way to guarantee the existence of an equilibrium
    is by requiring \(d\) to be a right-continuous function on \(E\),
    as in \textcite{Zapechelnyuk:2020}.
}
For each disclosure policy \(d\in\mathcal{D}\),
let \(\pi_d(\theta)\) and \(\gamma_d(\theta)\)
represent type \(\theta\)'s
maximized profit and equilibrium emission level, respectively.\footnote{
    If more than one emission levels maximize the firm's profit,
    then I assume the lowest one is implemented.
    This tie-breaking rule
    is in line with the standard assumption in the literature on mechanism design that
    the designer has the capability of selecting her preferred equilibrium in case of multiple equilibria.
}

\subsection{Implementability}

The Revelation Principle suggests that,
instead of considering all disclosure policies,
it is without loss of generality to focus directly on the set of emission schemes that could arise in equilibrium.
I say that a disclosure policy \(d\) \emph{implements} an emission scheme \(\mathbf{e}:\Theta\to E\)
if given \(d\), \((\mathbf{e}, \tilde{\mathbf{e}})\) is an equilibrium for some \(\tilde{\mathbf{e}}:E\to E\).
An emission scheme \(\mathbf{e}:\Theta\to E\) is \emph{implementable}
if there exists a disclosure policy \(d\in\mathcal{D}\) that implements \(\mathbf{e}\).

The following lemma establishes that
implementable emission schemes
can be characterized by the following standard constraints.\footnote{
    When the firm's profit function satisfies a single-crossing property,
    a well-known alternative characterization
    is presented in Proposition 1 of
    \textcite{MelumadShibano:1991}.
}
\begin{lemma}\label{lem:char-IC-IR}
    An emission scheme \(\mathbf{e}:\Theta\to E\) is implementable
    if and only if
    \begin{align}
        \pi(\theta, \mathbf{e}(\theta)) &\geq \pi(\theta, \mathbf{e}(\theta')), &\text{for all }\theta,\theta'\in\Theta, \tag{IC} \label{eq:IC}\\
        \text{and}\quad \pi(\theta, \mathbf{e}(\theta)) &\geq \pi(\theta, \bar{e}), &\text{for all } \theta\in\Theta. \tag{IR} \label{eq:IR}
    \end{align}
\end{lemma}
\begin{proof}
    See the Appendix.
\end{proof}
The proof of the ``only if'' part is straightforward.
If inequality~\eqref{eq:IC} is violated,
then the firm would deviate to the emission level \(\mathbf{e}(\theta')\),
which is belief-compatible given that it is chosen by some other types in equilibrium.
If inequality~\eqref{eq:IR} is violated,
then the firm would be better off by adopting the highest possible emission level \(\bar{e}\),
which serves as an outside option for all types of firms.
To prove the ``if'' part,
I show
that an emission scheme that satisfies these two constraints
can be implemented by a disclosure policy
which fully discloses the emission levels that are supposed to be chosen by some types,
and pools all the other emissions to higher levels so as to prevent them from being chosen by any type.

\section{The Value of Transparency}

In this section,
I compare the role of disclosure transparency across different informational environments.
To that end,
it is useful to introduce the following partial order over \(\mathcal{D}\).
\begin{definition}[Transparency]
    Given two distinct disclosure policies \(d,d'\in\mathcal{D}\),
    I say that \(d\) is more transparent than \(d'\) if the partition associated with \(d\) is finer than the one associated with \(d'\).\footnote{
        Given partitions \(P_1, P_2\) of a set \(X\),
        partition \(P_1\) is said to be \emph{finer} than \(P_2\) if every element in \(P_1\) is a subset of some element in \(P_2\).
    }
\end{definition}
\subsection{First Best}

I first consider the setting in which there is no information asymmetry between the firm and the regulator,
as if the regulator observes the firm's type
before committing to a disclosure policy.
Consequently,
dropping the incentive compatibility constraint in Lemma~\ref{lem:char-IC-IR} 
immediately gives the following proposition.
\begin{proposition}
    \label{prop:first-best}
    Let
    \(
        \underline{\mathbf{e}}(\theta) \coloneqq \min\{e\in E : \pi(\theta, e) \geq \pi(\theta, \bar{e})\}.
    \)
    Without information asymmetry,
    an emission level \(e^*\in E\) can be implemented
    if and only if \(e^* \geq \underline{\mathbf{e}}(\theta)\).
    Such an emission level can be implemented by the following binary disclosure:
    \[
        d(e) = \begin{cases}
            e^*, & \text{if }e\leq e^*,\\
            d(e) = \bar{e}, & \text{otherwise}.
        \end{cases}
    \]
\end{proposition}

Proposition~\ref{prop:first-best} demonstrates that,
when there is no information asymmetry,
the least transparent \emph{informative} disclosure policy---binary disclosure---is sufficient to achieve the first best.
More transparent policies do not improve welfare any further.
The intuition is straightforward.
Without information asymmetry,
the only constraint facing the regulator is inequality~\eqref{eq:IR}.
This individual rationality constraint arises from the regulator's limited authority,
as she can neither enforce the adoption of her most preferred emission level,
nor discipline the firm beyond 
exposing the firm's emissions through
disclosure regulation.
Indeed, the emission levels below \(\underline{\mathbf{e}}(\theta)\) can not
even be implemented when emissions are observable and verifiable.
Other than that,
the regulator is able to tackle the moral hazard problem to the maximum extent:
any emission level above \(\underline{\mathbf{e}}(\theta)\)
is implementable.
To implement \(e^*\in[\underline{\mathbf{e}}(\theta), \bar{e}]\),
the regulator adopts the binary disclosure that pools together all emissions weakly below,
and pools together all emissions strictly above.
The firm would then choose \(e^*\),
as deviating to a lower emission
would decrease the revenue without saving any cost of capital,
whereas deviating to a higher emission 
would yields a profit of \(\pi(\theta,\bar{e})\) at best,
and thus will not be profitable either.

\begin{figure}[htb]
    \centering
    \begin{subfigure}[b]{0.48\textwidth}
        \centering
        \includegraphics[width=\textwidth]{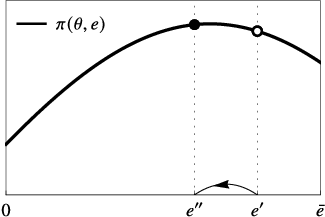}
        \caption{Transparency decreases emission}
        \label{fig:transparency-decreases-emission}
    \end{subfigure}
    \hfill
    \begin{subfigure}[b]{0.48\textwidth}
        \centering
        \includegraphics[width=\textwidth]{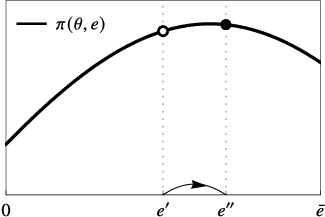}
        \caption{Transparency increases emission}
        \label{fig:transparency-increases-emission}
    \end{subfigure}

       \caption{
           Both figures illustrate the effect of increasing transparency
           from a disclosure policy represented by the partition \(\{[0,e'],(e',\bar{e}]\}\).
           This policy induces an equilibrium emission level \(e'\),
           as it yields a higher profit than the only other belief-compatible emission level \(\bar{e}\).
           In Figure~\ref{fig:transparency-decreases-emission},
           transparency is increased by breaking \([0, e']\)
           into intervals \([0,e'']\) and \((e'',e']\),
           which leads to a lower equilibrium emission level \(e''\).
           In Figure~\ref{fig:transparency-increases-emission},
           transparency is increased by breaking \((e', \bar{e}]\)
           into intervals \((e',e'']\) and \((e'',\bar{e}]\),
           which leads a higher equilibrium emission level \(e''\).
        }
       \label{fig:transparency-emission}
\end{figure}

The opaque structure in the binary disclosure
reveals that transparency does not necessarily reduce emissions,
which may seem counterintuitive at first glance.
To understand this claim,
observe that the game can be effectively decomposed into two stages.
First, the regulator chooses a disclosure policy \(d\in\mathcal{D}\),
which amounts to offering the firm a set of belief-compatible emission levels \(\tilde{E}_d\).
Second, the firm chooses an emission level from \(\tilde{E}_d\) to maximize its profit.
A more transparent policy
produces a larger set of belief-compatible emission levels,
potentially inducing the firm to switch to a level
that could not previously be sustained in equilibrium.
This newly selected equilibrium emission could be lower than the previous one,
as depicted in Figure~\ref{fig:transparency-decreases-emission},
or might as well be higher than the previous emission level,
as illustrated in Figure~\ref{fig:transparency-increases-emission}.
In this latter scenario,
increased disclosure transparency
leads to higher emissions,
thereby manifesting a trade-off between
transparency and externality.

In terms of the firm's profit,
the above argument reveals an even more striking fact:
increased transparency never makes the firm worse off.
The reason is simply that a more transparent disclosure policy enlarges the set of belief-compatible emission levels,
which serves as the feasible set of the firm's optimization problem.
Intuitively,
disclosure policies facilitate monitoring,
and more transparent policies enable a broader range of emission levels to be communicated to the market.
Unsurprisingly,
the firm would then opt for the one in its favor,
rather than being concerned about the resulting externality.
This observation can be formally summarized as follows.
\begin{lemma}
    \label{lem:transparent-profit}
    If \(d\) is more transparent than \(d'\), then \(\pi_d(\theta) \geq \pi_{d'}(\theta)\) for all \(\theta\in\Theta\).
\end{lemma}
\begin{proof}
    Since \(d\) is more transparent than \(d'\),
    we have \(\tilde{E}_{d'} \subset \tilde{E}_d\),
    which implies \(\gamma_{d'}(\theta) \in \tilde{E}_d\)
    and thus the emission level \(\gamma_{d'}(\theta)\)
    is belief-compatible under policy \(d\).
    Then we must have \(\pi_d(\theta) = \pi(\theta, \gamma_{d}(\theta)) \geq \pi(\theta, \gamma_{d'}(\theta)) = \pi_{d'}(\theta)\).
    Because if this is not the case,
    the firm would receive a higher profit by deviating from \(\gamma_{d}(\theta)\) to \(\gamma_{d'}(\theta)\) under policy \(d\).
\end{proof}

\subsection{Second Best}
Now,
I return to the initial setting in which the firm's type is privately known
and not observable to the regulator.
The uncertainty of \(\theta\)
make it challenging for the regulator
to identify a single level of emission
that would arise in equilibrium.
As a result,
under asymmetric information
the regulator may favor greater transparency over
the binary disclosure discussed earlier.

Increasing transparency tends to raise the firm's expected profit,
as it holds type by type
by Lemma~\ref{lem:transparent-profit}.
This result may be considered as an argument in favor of transparency, provided the regulator values the firm's private benefit along with emission reduction.
Nonetheless,
a more fundamental rationale for transparency
lies in its impact on externalities under adverse selection.
By choosing a disclosure policy,
the regulator essentially offers the firm a menu of belief-compatible emission levels,
with the maximal level \(\bar{e}\)
included by default so as to respect the individual rationality constraint.
Depending on the firm's type distribution,
the regulator might want to offer a richer menu,
by employing a more transparent policy,
to screen the firm's type more effectively.
In light of this,
increased transparency might be preferable for internalizing externalities,
largely because it mitigates adverse selection,
rather than moral hazard as suggested by conventional wisdom.

Nevertheless,
transparency might as well exacerbate adverse selection
through the transparency versus externality trade-off.
While a more transparent disclosure policy may induce certain types of firms to lower their emissions,
it might still raise the expected emissions,
as some other types
find it profitable to switch to
higher levels of belief-compatible emissions
made available by the increased transparency.
Overall,
the effect of transparency on emissions depends on the firm's type distribution,
as well as the region where transparency is increased.
The non-monotonic relationship between transparency and emission
discussed earlier
continues to hold under asymmetric information.

\section{Transparency and Welfare}

Having examined the impact of transparency on emissions and the firm's profit,
now I turn to its welfare implications.

Given a disclosure policy \(d\in\mathcal{D}\),
denote by \(\Pi(d) \coloneqq \int_\Theta \pi_d(\theta)\dd{F(\theta)}\) the expected profit of the firm,
and by \(\Gamma(d)\coloneqq \int_\Theta \gamma_d(\theta)\dd{F(\theta)}\) the expected emission level in the equilibrium.

\begin{figure}[htb]
    \centering
    \includegraphics[width=0.95\textwidth]{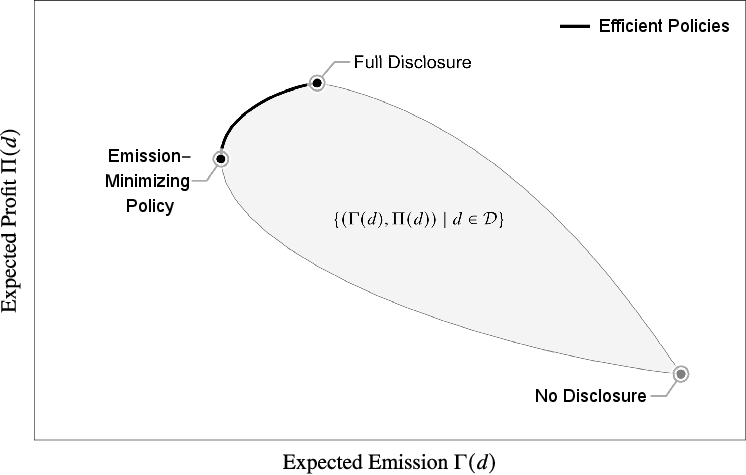}
    \caption{
        The emission-profit possibility set
        and the efficient frontier.
    }
    \label{fig:possibility-set}

\end{figure}

\begin{definition}[Efficient Policy]
    A disclosure policy \(d\) is \emph{Pareto efficient} if there is no disclosure policy \(d'\in\mathcal{D}\) such that
    \(\Pi(d') \geq \Pi(d)\) and \(\Gamma(d') \leq \Gamma(d)\),
    with at least one inequality strict.
\end{definition}
Figure~\ref{fig:possibility-set} depicts a typical set of feasible emission-profit pairs \((\Gamma(d), \Pi(d))\),
with the solid boundary corresponding to its Pareto frontier.

The uninformative disclosure,
or equivalently no disclosure,
is clearly not efficient.
In fact, it can be viewed as the least efficient policy
in the following sense.
\begin{proposition}[No Disclosure]
    \label{prop:no-disclosure}
    The uninformative disclosure policy induces the highest emission level
and the lowest profit among all disclosure policies.
\end{proposition}
\begin{proof}
    The uninformative disclosure, i.e., \(d(e) = \bar{e}\) for all \(e\in E\),
    induces the emission level \(\bar{e}\) for all \(\theta\in\Theta\).
    Therefore \(\Gamma(d) = \bar{e} \geq \Gamma(d')\) for all \(d'\in\mathcal{D}\).
    The claim on the expected profit follows from inequality~\eqref{eq:IR} in Lemma~\ref{lem:char-IC-IR}.
\end{proof}

The uninformative disclosure can be thought of as the least transparent policy.
On the other end of the spectrum,
we have the most transparent disclosure policy
defined as follows.
\begin{definition}[Full Disclsoure]
    A disclosure policy \(d\) is a \emph{full-disclosure policy} if
    \(d(e) = e\) for all \(e\in\ehat(\Theta)\),
    where \(\ehat(\theta)\coloneqq \argmax_{e\in E} \pi(\theta, e)\).\footnote{
        Here, \(\ehat(\Theta)\) denotes the image of \(\ehat\).
        My definition of full disclosure is outcome-equivalent to the alternative that requires \(d(e) = e\) for all \(e\in E\).
    }
\end{definition}
Since the most desirable emission levels become observable to the market for each \(\theta\),
under full disclosure
all types of firm can attain
its highest possible equilibrium profit by choosing its most preferred emission level \(\ehat(\theta)\).

As the profit is maximized type by type under full disclosure,
we are able to conclude the following.
\begin{proposition}\label{prop:full-disclosure-profit-maximization}
    A full-disclosure policy is efficient,
    and induces the highest expected profit among all disclosure policies.
\end{proposition}

\begin{proof}
    The claim on the expected profit follows directly from Lemma~\ref{lem:transparent-profit}.
    Efficiency then follows from the strict concavity of \(\pi(\theta,\,\cdot\,)\),
    which implies there does not exist a disclosure policy that attains a lower expected emission without lowering the expected profit.
\end{proof}

Given the non-monotonic relationship
between transparency and emission as a result of
the transparency versus externality trade-off,
should the regulator seek more transparent disclosure policies
to combat carbon emissions?
The next proposition provides a partial answer. 

\begin{proposition}
    \label{prop:transparency-efficiency}
    If \(d'\) is more transparent than \(d\)
    and \(d\) is efficient,
    then \(\Gamma(d') \geq \Gamma(d)\).
\end{proposition}
\begin{proof}
    Given that \(d'\) is more transparent than \(d\),
    Lemma~\ref{lem:transparent-profit} implies that \(\Pi(d') \geq \Pi(d)\).
    If \(\Gamma(d') < \Gamma(d)\),
    then by definition \(d\) is not efficient,
    which is a contradiction.
\end{proof}

Starting with no disclosure,
more transparent disclosures improve efficiency
by mitigating the moral hazard problem.
As long as inefficiency remains,
the firm should embrace, rather than resist,
the emission reduction induced by increased transparency
because of the higher profit attained.
Nonetheless, 
since more transparent disclosures
tend to allocate the surplus to the firm,
transparency exacerbates the externality once efficiency is achieved.
Proposition~\ref{prop:transparency-efficiency}
asserts that as long as efficiency is achieved, attempting to further reduce emissions by pursuing transparency would be in vain.
This insight naturally leads to the following counterintuitive statement.

\begin{proposition}
    \label{prop:full-disclosure-pareto}
    A full-disclosure policy induces the highest expected emission level among all efficient disclosure policies.
\end{proposition}

\begin{proof}
    This proposition follows directly from 
    Proposition~\ref{prop:transparency-efficiency}.
\end{proof} 

Proposition~\ref{prop:full-disclosure-pareto}
makes clear what would be achieved when pursuing maximal transparency.
We learn from
Lemma~\ref{lem:transparent-profit}
and Proposition~\ref{prop:full-disclosure-profit-maximization} that full transparency maximizes the profit of the firm and therefore restores efficiency.
However,
when efficiency is considered as a precondition for disclosure policies,
perhaps surprisingly,
full transparency proves to be the worst in terms of internalizing externalities.

\section{Efficient Disclosure Policies}

\label{sec:optimal-policy}

In principle, an efficient disclosure policy could exhibit complex structures,
dividing the emission space into a mixture of multiple transparent and opaque regions.
Yet,
efficient disclosure policies can be simple.
Theorem~\ref{thm:frontier-char} in this section shows that
when focusing on a specific class of profit functions and type distributions,
any emission-profit pair on the Pareto frontier
can be attained by policies with a threshold structure.

The profit functions considered in this section
are assumed to take the following form.
\begin{assumption}
    \label{assump:linear-delegation}
    The profit function of the firm
    \(\pi(\theta, e) = \tilde{\pi}(\theta, e, \tilde{e})\)
    satisfies
    \[
        \pi(\theta, e) = \pi_0(e) - \theta \cdot (a e + b) ,
    \]
    for some some function \(\pi_0\) and constants \(a,b\in\mathbb{R}\).
\end{assumption}
For \(a > 0\),
a firm with a higher \(\theta\) behaves more environmentally friendly,
as its marginal private benefit of emission,
\(\pi_0'(e) - a \theta\),
is lower.

\begin{example}
    \label{ex:cost-of-capital}
    Consider a firm with a profit function
    \(\tilde{\pi}(\theta, e, \tilde{e}) = r(e) - \theta (a\tilde{e} + b)\),
    where \(\theta, a, b > 0\).
    Here,
    \(r(e)\) denotes the firm's revenue,
    which increases with the quantity produced
    and, thus, with emissions as well. 
    The term \(a\tilde{e} + b\) represents the cost of capital,
    which increases linearly with \(\tilde{e}\).
    The firm's private type \(\theta\) denotes the amount of capital
    that needs to be financed externally.
    Therefore, a firm with a higher \(\theta\) has greater financial needs,
    and therefore a stronger incentive to reduce emissions.
\end{example}

\begin{example}
    The profit function used in Example~\ref{ex:cost-of-capital}
    can also be interpreted in the context of a manager with ESG reputation concerns.
    Now, the term \(\theta(a\tilde{e} + b)\) becomes the reputation damage
    associated with the perceived level of emissions \(\tilde{e}\),
    and therefore managers with a higher \(\theta\) 
    have greater reputation concerns.
\end{example}

\begin{example}
    Consider the profit function
    \(\tilde{\pi}(\theta, e, \tilde{e}) = R - c(\tilde{e}) -  \abs{\theta}(\bar{e} - e)\) with \(\theta < 0\),
    where \(R\) is a constant representing the revenue created from the project,
    and \(c(\tilde{e})\) represents the cost of capital.
    The firm is equipped with a technology that allows it to reduce the emission from \(\bar{e}\) to \(e\)
    at a unit cost of
    \(\abs{\theta}\).
    Thus, firms with higher (less negative) \(\theta\) are more efficient at reducing emissions.
\end{example}

\begin{example}
    A firm (or its shareholders) with an intrinsic preference for low emissions
    might have a profit function
    \(\tilde{\pi}(\theta, e, \tilde{e}) = r(e) - c(\tilde{e}) - \theta a e\)
    where \(\theta, a > 0\).
    With \(r(\,\cdot\,)\) and \(c(\,\cdot\,)\) being interpreted in the same way as in the previous examples,
    the term
    \(\theta a e \) here represents the disutility from emissions.
    Thus, a higher \(\theta\)
    corresponds to a more socially responsible firm.
\end{example}

\begin{figure}[htb]
    \centering
    \includegraphics[width=0.65\textwidth]{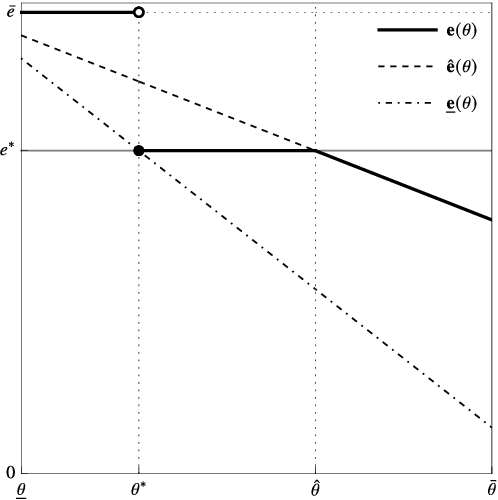}
    \caption{
        The emission scheme \(\mathbf{e}(\theta)\) induced by a threshold policy with threshold \(e^*\).
    }
    \label{fig:threshold-policy}
\end{figure}

Next, we focus on disclosure policies with the following threshold structure.
\begin{definition}[Threshold Disclosure Policies]
    A disclosure policy \(d\) is said to be a \emph{threshold policy}
    with threshold \(e^*\in E\)
    if \(d(e) = e\) for \(e \leq e^*\)
    and \(d(e) = \bar{e}\) otherwise.
\end{definition}
A threshold policy thus contains a transparent region at the bottom
and an opaque region at the top.
Figure~\ref{fig:threshold-policy} illustrates an example of the emission scheme implemented by a threshold policy,
with the constant \(a\) in Assumption~\ref{assump:linear-delegation}
being positive.\footnote{
    For \(a < 0\),
    the types in \(\Theta\) will be arranged in the opposite order in the following discussion.
    A higher \(\theta\) would then correspond to a firm that behaves less environmentally friendly.
}
Firms with type \(\theta\in[\underline{\theta}, \theta^*)\)
have \(\pi(\theta, e^*) < \pi(\theta, \bar{e})\)
and thus do not find it worthwhile to reduce emissions from \(\bar{e}\).
Firms with type \(\theta\in[\theta^*, \hat{\theta}]\) find \(\pi(\theta, e^*)\geq \pi(\theta, \bar{e})\).
They are induced to lower their emission to the threshold \(e^*\)
since all higher levels are pooled to \(\bar{e}\).
Lastly, firms with type \(\theta \in [\hat{\theta},\bar{\theta}]\)
have \(\ehat(\theta)\) in the transparent region,
and thus 
obtain the the highest possible profit by choosing \(\ehat(\theta)\).

\begin{theorem}[Pareto Frontier Characterization]
    \label{thm:frontier-char}
    Suppose the profit function of the firm satisfies
    Assumption~\ref{assump:linear-delegation}.
    If 
    \(f\) is continuously differentiable
    and
    \(\ln(f(\theta))'' \leq 0\) on \(\interior\Theta\),
    then for any efficient disclosure policy,
    there exists a threshold policy that induces the same expected emission and expected profit.
    Furthermore,
    if \(\ln(f(\theta))'' < 0\) on \(\interior\Theta\),
    then the corresponding threshold is unique in \([0, \max\ehat(\Theta)]\).\footnote{
        Recall that any threshold policies with a threshold weakly larger than \(\max\ehat(\Theta)\)
        induce the same emission scheme, namely \(\ehat\).
        Thus, when \(\max\ehat(\Theta) < \bar{e}\),
        the scope within which the threshold remains unique cannot be extended beyond \(\max\ehat(\Theta)\).
    } 
\end{theorem}
\begin{proof}
    See the Appendix.
\end{proof}

Many commonly used probability distributions are log-concave.
Theorem~\ref{thm:frontier-char} asserts that,
with such type distributions
and the profit functions outlined above,
we can restrict attention to threshold policies without loss.
Consequently,
provided that the regulator adopts a welfare function
favoring lower expected emissions and higher expected profits,
the search among all disclosure policies
amounts to a straightforward single-variable optimization problem.

We know from Proposition~\ref{prop:full-disclosure-pareto} that
the emission-profit pair induced by full disclosure
must lie on the Pareto frontier.
However,
the Pareto frontier may contain no other points.
The following proposition provides sufficient conditions for this situation.

\begin{proposition}
    \label{prop:full-disclosure}
    Suppose the profit function of the firm satisfies
    Assumption~\ref{assump:linear-delegation} with \(a > 0\) (resp.\ \(a < 0\)),
    and
    \(f\) is continuously differentiable
    on \(\interior\Theta\).
    If there exists some \(\theta_\blacktriangle\in\Theta\) such that
    \begin{enumerate}[label=(\roman*)]
        \item 
        \label{cond:full-disclosure-abandoned}
        \(\underline{\mathbf{e}}(\theta) = \bar{e}\) 
        on \([\underline{\theta},\theta_\blacktriangle]\)
        (resp.\ on \([\theta_\blacktriangle, \bar{\theta}]\)),
        \item 
        and \(f\) is nonincreasing on \([\theta_\blacktriangle, \bar{\theta}]\)
        (resp.\ nondecreasing on \([\underline{\theta},\theta_\blacktriangle]\)),
    \end{enumerate}
    then full disclosure is the only efficient policy.
\end{proposition}

Note that
firms with \(\underline{\mathbf{e}}(\theta) = \bar{e}\)
are irresponsive to disclosure policies.
They will always choose the level \(\bar{e}\) in any equilibrium.
Thus,
the conditions specified in Proposition~\ref{prop:full-disclosure}
describe an economy 
where market forces are relatively weak in disciplining firms,
with only the types at the tail of the distribution being responsive to regulation.
In such a scenario,
no room remains for the regulator to lower the emissions further through the design of disclosure policies,
making full disclosure the only sensible choice.

\appendix
\appendixpage

Section~\ref{sec:proof-lem-global-sufficiency} and~\ref{sec:proof-thm-frontier-char}
in this appendix
present the proof of Theorem~\ref{thm:frontier-char}.
Instead of adopting the profit function
\(\pi(\theta, e) = \pi_0(e) - \theta(a e + b)\)
as specified in Assumption~\ref{assump:linear-delegation},
in this appendix I assume \(a=-1\) and \(b=0\),
which yields
\(\pi(\theta, e) = \pi_0(e) + \theta e\).
Once Theorem~\ref{thm:frontier-char} is proven for this specific case,
the results extend immediately to any \(a,b\in \mathbb{R}\).
This is because multiplying \(\theta\) by a nonzero constant
merely rescales \(\theta\)
without affecting the log-concavity of the type distributions,\footnote{
    Theorem~\ref{thm:frontier-char} is trivial if \(a=0\).
}
whereas subtracting the term \(\theta b\) from the profit function for each \(\theta\) does not alter the equilibrium emission scheme for any given disclosure policy.

Note that
when the constant \(a\) is negative,
contrary to Figure~\ref{fig:threshold-policy},
both functions \(\ehat(\,\cdot\,)\) and \(\underline{\mathbf{e}}(\,\cdot\,)\) are strictly increasing,
resulting in \(\hat{\theta} \leq \theta^*\).
Consequently, for \(a<0\),
the equilibrium emission scheme \(\mathbf{e}\) induced by a threshold policy with a threshold \(e^*\) can be written as:
\[
    \mathbf{e}(\theta) = \begin{cases}
        \ehat(\theta); & \theta \leq \hat{\theta},\\
        e^*; & \theta\in (\hat{\theta}, \theta^*],\\
        \bar{e}; & \theta \in (\theta^*, \bar{\theta}],
    \end{cases}
\]
where
\begin{align}
    \label{eq:theta-hat}
    \hat{\theta} \coloneqq \begin{cases}
        \bar{\theta}; &\text{if }e^*>\ehat(\bar{\theta}),\\
        \ehat^{-1}(e^*); &\text{if } e^*\in[\ehat(\underline{\theta}), \ehat(\bar{\theta})],\\
        \underline{\theta}; &\text{if }e^*<\ehat(\underline{\theta}),
    \end{cases}
\end{align}
and
\begin{align}
    \label{eq:theta-star}
    \theta^* \coloneqq \begin{cases}
        \bar{\theta}; &\text{if }e^*>\underline{\mathbf{e}}(\bar{\theta}),\\
        \underline{\mathbf{e}}^{-1}(e^*); &\text{if } e^*\in[\underline{\mathbf{e}}(\underline{\theta}), \underline{\mathbf{e}}(\bar{\theta})],\\
        \underline{\theta}; &\text{if }e^*<\underline{\mathbf{e}}(\underline{\theta}).
    \end{cases}
\end{align}

In the remainder of this appendix,
\(\hat{\theta}\) and \(\theta^*\)
refer to the functions defined above,
whereas the explicit dependency on \(e^*\) is omitted when clear from the context.

\section{Optimality of Threshold Policies}
\label{sec:proof-lem-global-sufficiency}
In this section,
I provide the proof
of Lemma~\ref{lem:global-sufficiency} below,
which establishes a sufficient condition
for the optimality of a threshold policy
when the regulator maximizes 
a weighted sum of the public benefit \(-\Gamma\) and the private benefit \(\Pi\).

Let \(\mathcal{D}^*\) denote the set of all threshold policies,
and
consider the optimization problems
\begin{align}
    \max_{d\in\mathcal{D}} \{- \alpha \Gamma(d) + (1-\alpha) \Pi(d)\}, \tag{\(\mathcal{P}_\alpha\)}
    \label{prob:all-policies}
\end{align}
and
\begin{align}
    \max_{d\in\mathcal{D}^*} \{- \alpha \Gamma(d) + (1-\alpha) \Pi(d)\}, \tag{\(\mathcal{P}_\alpha^*\)}
    \label{prob:threshold-policies}
\end{align}
where \(\alpha\in[0,1]\).

In the remainder of this Appendix,
let \(W:E\to\mathbb{R}\) denote the function that maps a threshold \(e^*\in E\)
of a threshold policy \(d\in\mathcal{D}^*\)
to the objective value 
\(- \alpha \Gamma(d) + (1-\alpha) \Pi(d)\).

\begin{lemma}
    
    \label{lem:global-sufficiency}
    Suppose \((1-\alpha)F(\theta) + \alpha f(\theta)\) is quasiconcave on \(\Theta\).
    A threshold policy with a threshold \(e^*\)
    solves Problem~\eqref{prob:all-policies}
    if it solves Problem~\eqref{prob:threshold-policies}.

\end{lemma}

\begin{proof}

First, Problem~\eqref{prob:all-policies} can be rewritten as\footnote{See, e.g., \textcite{Amador-Bagwell:2022}.}
\begin{align*}
    \max_{\mathbf{e}} &\int_{\underline{\theta}}^{\bar{\theta}} [-\alpha \mathbf{e}(\theta) + (1-\alpha) \pi(\theta, \mathbf{e}(\theta))] \dd{F(\theta)} 
    \quad\text{subject to:}\\
    \quad &\pi(\theta, \mathbf{e}(\theta)) + \int_\theta^{\bar{\theta}} \mathbf{e}(\tilde{\theta})\dd{\tilde{\theta}} = \pi(\bar{\theta}, \mathbf{e}(\bar{\theta})),
    \quad\text{for all } \theta\in\Theta,\\
    \quad &\pi(\theta, \mathbf{e}(\theta)) \geq \pi(\theta,\bar{e}),
    \quad\text{for all } \theta\in\Theta,\\
    \quad &\mathbf{e}:\Theta\to E \text{ nondecreasing}.
\end{align*}

Next, I follow \textcite{Amador-Bagwell:2013}
and write the incentive constraint as two inequalities:
\begin{align*}
    -\int_\theta^{\bar{\theta}} \mathbf{e}(\tilde{\theta})\dd{\tilde{\theta}}
    -\pi(\theta, \mathbf{e}(\theta))
    +\pi(\bar{\theta}, \mathbf{e}(\bar{\theta})) &\leq 0
    \quad\text{for all } \theta\in\Theta,\\
    \int_\theta^{\bar{\theta}} \mathbf{e}(\tilde{\theta})\dd{\tilde{\theta}}
    +\pi(\theta, \mathbf{e}(\theta))
    -\pi(\bar{\theta}, \mathbf{e}(\bar{\theta})) &\leq 0
    \quad\text{for all } \theta\in\Theta.
\end{align*}
With the choice set for \(\mathbf{e}\)
defined as \(\Phi\coloneqq \{\mathbf{e} \mid \mathbf{e}:\Theta\to E,\text{ and } \mathbf{e}\text{ nondecreasing}\}\),
Problem~\eqref{prob:all-policies} can further be written as
\begin{align}
    \max_{\mathbf{e}\in\Phi} &\int_{\underline{\theta}}^{\bar{\theta}} [-\alpha \mathbf{e}(\theta) + (1-\alpha) \pi(\theta, \mathbf{e}(\theta))] \dd{F(\theta)} 
    \quad\text{subject to:} \nonumber\\
    \label{ineq:IC-1}
    &-\int_\theta^{\bar{\theta}} \mathbf{e}(\tilde{\theta})\dd{\tilde{\theta}}
    -\pi(\theta, \mathbf{e}(\theta))
    +\pi(\bar{\theta}, \mathbf{e}(\bar{\theta})) \leq 0
    \quad\text{for all } \theta\in\Theta,\\
    \label{ineq:IC-2}
    &\int_\theta^{\bar{\theta}} \mathbf{e}(\tilde{\theta})\dd{\tilde{\theta}}
    +\pi(\theta, \mathbf{e}(\theta))
    -\pi(\bar{\theta}, \mathbf{e}(\bar{\theta})) \leq 0
    \quad\text{for all } \theta\in\Theta,\\
    \label{ineq:IR}
    &-\pi(\theta, \mathbf{e}(\theta)) + \pi(\theta,\bar{e}) \leq 0
    \quad\text{for all } \theta\in\Theta.
\end{align}

Now, consider a threshold policy with a threshold \(e^*\) that solves Problem~\eqref{prob:threshold-policies}.
Denote by \(\mathbf{e}^*\) the resulting emission scheme.
Without loss of generality,
I assume
that \(e^* \leq \ehat(\bar{\theta})\),
which implies that
\(\hat{\theta} = \bar{\theta}\)
if and only if \(e^* = \ehat(\bar{\theta})\).\footnote{
    Note that,
    when \(\ehat(\bar{\theta}) < \bar{e}\),
    any threshold policy with \(e^*\in[\ehat(\bar{\theta}), \bar{e}]\)
    gives to the same emission scheme.
    Therefore,
    if a threshold policy with a threshold larger than \(\ehat(\bar{\theta})\)
    solves Problem~\eqref{prob:all-policies},
    so does a threshold policy with threshold \(e^* = \ehat(\bar{\theta})\).
}
In what follows,
I show that \(\mathbf{e}^*\) solves the optimization problem above.

Let \(\Lambda_1(\theta)\), \(\Lambda_2(\theta)\), and \(\Psi(\theta)\)
denote the (cumulative) multiplier functions associated with
inequalities~\eqref{ineq:IC-1},~\eqref{ineq:IC-2} and~\eqref{ineq:IR}, respectively.
These functions are restricted to be nondecreasing on \(\Theta\).
Let \(\Lambda(\theta) \coloneqq \Lambda_1(\theta) - \Lambda_2(\theta)\).
The Lagrangian can be written as follows:
\begin{align*}
    \mathcal{L}(\mathbf{e}\mid \Lambda, \Psi) =
    &\int_{\underline{\theta}}^{\bar{\theta}} [-\alpha \mathbf{e}(\theta) + (1-\alpha) \pi(\theta, \mathbf{e}(\theta))] \dd{F(\theta)} \\
    &+ \int_{\underline{\theta}}^{\bar{\theta}}
    \left(
        \int_{\theta}^{\bar{\theta}}
        \mathbf{e}(\tilde{\theta})\dd{\tilde{\theta}}
        +\pi(\theta, \mathbf{e}(\theta))
        -\pi(\bar{\theta}, \mathbf{e}(\bar{\theta}))
    \right)
    \dd{\Lambda(\theta)}\\
    &+ \int_{\underline{\theta}}^{\bar{\theta}}
    (\pi(\theta, \mathbf{e}(\theta)) - \pi(\theta, \bar{e})) \dd{\Psi(\theta)}.
\end{align*}
Integration by parts gives
\begin{align*}
    \mathcal{L}(\mathbf{e}\mid \Lambda, \Psi) =&
    \int_{\underline{\theta}}^{\bar{\theta}} -\alpha \mathbf{e}(\theta) \dd{F(\theta)}
    +\int_{\underline{\theta}}^{\bar{\theta}} \pi(\theta, \mathbf{e}(\theta)) \dd{((1-\alpha)F(\theta) + \Lambda(\theta) + \Psi(\theta))}\\
    &+ \int_{\underline{\theta}}^{\bar{\theta}} (\Lambda(\theta) -  \Lambda(\underline{\theta})) \mathbf{e}(\theta) \dd{\theta}
    - \pi(\bar{\theta}, \mathbf{e}(\bar{\theta}))(\Lambda(\bar{\theta}) - \Lambda(\underline{\theta}))\\
    &- \int_{\underline{\theta}}^{\bar{\theta}} \pi(\theta,\bar{e})\dd{\Psi(\theta)}.
\end{align*}

I propose the following multipliers:
\[
    \Lambda(\theta) \coloneqq \begin{cases}
        0, & \theta = \underline{\theta},\\
        \alpha f(\theta), & \theta\in(\underline{\theta}, \hat{\theta}),\\
        (1-\alpha)F(\theta^*) + \alpha f(\theta^*) - (1-\alpha)F(\theta), &
        \theta\in[\hat{\theta}, \theta^*)\backslash\{\underline{\theta}\},
        \\
        \alpha f(\theta), & 
        \theta\in[\theta^*, \bar{\theta}),
        \\
        \alpha f(\theta^*) - A, & \theta = \bar{\theta},
    \end{cases}
\]
and
\[
    \Psi(\theta) \coloneqq \begin{cases}
        0, & \theta\in[\underline{\theta}, \theta^*),\\
        (1-\alpha)F(\theta^*) + \alpha f(\theta^*) - (1-\alpha)F(\theta) - \alpha f(\theta), &
        \theta \in [\theta^*, \bar{\theta}),
        \\
        (1-\alpha) F(\theta^*) - (1-\alpha) F(\bar{\theta}) + A, & \theta =\bar{\theta},
    \end{cases}
\]
where
\[
    A \coloneqq \begin{cases}
        \frac{-1}{\pi_e(\theta^*, e^*)}
        \int_{\hat{\theta}}^{\theta^*}
            [-\alpha + (1-\alpha) \pi_{e}(\theta, e^*)]
        \dd{F(\theta)}, & \text{if } \pi_e(\theta^*, e^*)\neq 0\\
        0, & \text{otherwise}.
    \end{cases}
\]

Note that
these functions are well defined
even if
\(\underline{\theta}=\hat{\theta}\)
or \(\theta^* = \bar{\theta}\).
Also,
note that we must have
\(e^* > \underline{\mathbf{e}}(\underline{\theta})\),
which implies \(\theta^* > \underline{\theta}\).
Because
if otherwise,
the induced emission scheme becomes \(\mathbf{e}^*(\theta) = \bar{e}\)
for all \(\theta\in\Theta\),
which is clearly not optimal within the class of threshold policies.

By Lemma~\ref{lem:foc},
we have
\(A = \alpha f(\theta^*)\) if \(\theta^* < \bar{\theta}\),
and \(A\in[0,\alpha f(\theta^*)]\) otherwise.
Therefore,
if \(\Psi(\theta)\) is discontinuous at \(\bar{\theta}-\),
then its jump at \(\bar{\theta}\) cannot be negative.
Moreover,
since the function \((1-\alpha)F(\theta) + \alpha f(\theta)\) is quasiconcave,
Lemma~\ref{lem:peak-range}
implies that it is nonincreasing on \([\theta^*, \bar{\theta}]\).
Therefore, \(\Psi\) is nondecreasing on \(\Theta\).
By letting \(\Lambda_1(\theta) \coloneqq (1-\alpha)F(\theta) + \Lambda(\theta) + \Psi(\theta)\),
and \(\Lambda_2\coloneqq (1-\alpha) F(\theta) + \Psi(\theta)\),
we will see that
the function \(\Lambda\) can be written as the difference between two nondecreasing functions,
where the monotonicity of \(\Lambda_1\) will be verified shortly.

With \(\Lambda(\underline{\theta}) = 0\),
the Lagrangian becomes
\begin{align*}
    \mathcal{L}(\mathbf{e}\mid \Lambda, \Psi) =&
    \int_{\underline{\theta}}^{\bar{\theta}} -\alpha \mathbf{e}(\theta) \dd{F(\theta)}
    +\int_{\underline{\theta}}^{\bar{\theta}} \pi(\theta, \mathbf{e}(\theta)) \dd{((1-\alpha)F(\theta) + \Lambda(\theta) + \Psi(\theta))}\\
    &+ \int_{\underline{\theta}}^{\bar{\theta}} \Lambda(\theta) \mathbf{e}(\theta) \dd{\theta}
    -\int_{\underline{\theta}}^{\bar{\theta}} \pi(\theta,\bar{e})\dd{\Psi(\theta)}
    - \Lambda(\bar{\theta})\pi(\bar{\theta}, \mathbf{e}(\bar{\theta})).
\end{align*}

\paragraph*{Concavity of the Lagrangian.}

Since \(\pi(\theta, e)\)
is concave in \(e\),
the Lagrangian is concave in \(\mathbf{e}\)
if \((1-\alpha)F(\theta) + \Lambda(\theta) + \Psi(\theta)\)
is nondecreasing in \(\theta\).
With the proposed multipliers,
we have
\[
    (1-\alpha)F(\theta) + \Lambda(\theta) + \Psi(\theta)
    = \begin{cases}
        0, & \theta = \underline{\theta},\\
        (1-\alpha)F(\theta)
        +
        \alpha f(\theta),
        &\theta\in(\underline{\theta}, \hat{\theta}),\\
        (1-\alpha)F(\theta^*) + \alpha f(\theta^*),
        &\theta\in[\hat{\theta}, \bar{\theta}]\backslash\{\underline{\theta}\}.
    \end{cases}
\]
This expression is clearly nondecreasing
if \(\hat{\theta} = \underline{\theta}\).
Now suppose that \(\hat{\theta} > \underline{\theta}\).
As the function \((1-\alpha) F(\theta) + \alpha f(\theta)\)
is assumed to be quasiconcave,
Lemma~\ref{lem:peak-range} implies that
it is nondecreasing on \((\underline{\theta}, \hat{\theta})\).
Then, the above expression is nondecreasing on \(\Theta\) as long as
it has a nonnegative jump at \(\hat{\theta}\).
This is clearly the case if \(\hat{\theta} = \theta^* = \bar{\theta}\).
I will verify this fact later for the case of \(\underline{\theta} < \hat{\theta} < \bar{\theta}\).

\paragraph*{Maximizing the Lagrangian.}

I now show that the emission scheme \(\mathbf{e}^*\)
maximizes the Lagrangian.
I first extend \(\pi(\theta, \,\cdot\,)\)
to the entire positive ray of the real line
by defining 
\[
    \hat{\pi}(\theta, e)
    \coloneqq \begin{cases}
        \pi(\theta, e), &\text{for } e\in[0,\bar{e}],\\
        \pi(\theta, \bar{e}) + \pi_{e}(\theta, \bar{e})(e - \bar{e}), &\text{for } e\in(\bar{e}, \infty),
    \end{cases}
\]
for all \(\theta\in\Theta\).
Consequently,
the choice set \(\Phi\)
can be extended to a convex cone \(\hat{\Phi}\coloneqq \{\mathbf{e} \mid \mathbf{e}:\Theta\to \R_{+}, \text{ and } \mathbf{e} \text{ nondecreasing}\}\).
I then define the extended Lagrangian as
\begin{align*}
    \hat{\mathcal{L}}(\mathbf{e}\mid \Lambda, \Psi) =&
    \int_{\underline{\theta}}^{\bar{\theta}} -\alpha \mathbf{e}(\theta) \dd{F(\theta)}
    +\int_{\underline{\theta}}^{\bar{\theta}} \hat{\pi}(\theta, \mathbf{e}(\theta)) \dd{((1-\alpha)F(\theta) + \Lambda(\theta) + \Psi(\theta))}\\
    &+ \int_{\underline{\theta}}^{\bar{\theta}} \Lambda(\theta) \mathbf{e}(\theta) \dd{\theta}
    -\int_{\underline{\theta}}^{\bar{\theta}} \hat{\pi}(\theta,\bar{e})\dd{\Psi(\theta)}
    - \Lambda(\bar{\theta})\hat{\pi}(\bar{\theta}, \mathbf{e}(\bar{\theta})).
\end{align*}
For the same reason as before,
\(\hat{\mathcal{L}}\) is concave in \(\mathbf{e}\).
Moreover,
because \(\hat{\mathcal{L}}\) and \(\mathcal{L}\) coincide on \(\Phi\),
I can say that
if \(\mathbf{e}^*\) maximizes the extended Lagrangian \(\hat{\mathcal{L}}(\mathbf{e}\mid \Lambda, \Psi)\)
over \(\hat{\Phi}\),
then it also maximizes Lagrangian \(\mathcal{L}(\mathbf{e}\mid \Lambda, \Psi)\) over \(\Phi\).
According to Lemma A.2 in \textcite{Amador-Werning-Angeletos:2006},
I can then say that if
\begin{align*}
    &\partial\hat{\mathcal{L}}(\mathbf{e}^*; \mathbf{e}^*\mid \Lambda, \Psi) = 0, \\
    &\partial\hat{\mathcal{L}}(\mathbf{e}^*; \mathbf{x}\mid \Lambda, \Psi) \leq 0; \quad \text{for all } \mathbf{x}\in\hat{\Phi},
\end{align*}
then \(\mathbf{e}^*\) maximizes \(\hat{\mathcal{L}}(\mathbf{e}\mid \Lambda, \Psi)\) over \(\hat{\Phi}\).
In what follows, I show that these conditions are satisfied.

Taking the Gateaux differential in the direction \(\mathbf{x}\in\hat{\Phi}\),
we obtain
\begin{align*}
    \partial{\hat{\mathcal{L}}(\mathbf{e}^*; \mathbf{x}\mid \Lambda, \Psi)}
    =& \int_{\underline{\theta}}^{\bar{\theta}} \left[-\alpha f(\theta) + \Lambda(\theta)\right] \mathbf{x}(\theta)\dd{\theta}\\
    &+ \int_{\underline{\theta}}^{\bar{\theta}} \pi_{e}(\theta, \mathbf{e}^*(\theta)) \mathbf{x}(\theta)\dd{((1-\alpha)F(\theta) +\Lambda(\theta) + \Psi(\theta))}\\
    &-\Lambda(\bar{\theta})\pi_e(\bar{\theta}, \mathbf{e}^*(\bar{\theta}))\mathbf{x}(\bar{\theta}).
\end{align*}
Given our choice of \(\Lambda(\theta)\) within the interval \((\underline{\theta}, \hat{\theta})\)
and the observation that \(\pi_{e}(\theta, \mathbf{e}^*(\theta)) = \pi_{e}(\theta, \ehat(\theta)) = 0\)
for all \(\theta\in[\underline{\theta}, \hat{\theta}]\) whenever \(\hat{\theta} > \underline{\theta}\),
it follows that
\begin{align*}
    \partial{\hat{\mathcal{L}}(\mathbf{e}^*; \mathbf{x}\mid \Lambda, \Psi)}
    =&\int_{\hat{\theta}}^{\bar{\theta}} \left[-\alpha f(\theta) + \Lambda(\theta)\right] \mathbf{x}(\theta)\dd{\theta}\\
    &+ \int_{\hat{\theta}}^{\bar{\theta}} \pi_{e}(\theta, \mathbf{e}^*(\theta)) \mathbf{x}(\theta)\dd{((1-\alpha)F(\theta) +\Lambda(\theta) + \Psi(\theta))}\\
    &-\Lambda(\bar{\theta})\pi_e(\bar{\theta}, \mathbf{e}^*(\bar{\theta}))\mathbf{x}(\bar{\theta}).
\end{align*}

Integration by parts yields
\begin{align*}
    \partial{\hat{\mathcal{L}}(\mathbf{e}^*; \mathbf{x}\mid \Lambda, \Psi)}
    =&
    \left[
        \int_{\hat{\theta}}^{\bar{\theta}} \left[-\alpha f(\theta) + \Lambda(\theta) \right]\dd{\theta}
    \right.\\
    &\left.
        \quad +\int_{\hat{\theta}}^{\bar{\theta}} \pi_{e}(\theta, \mathbf{e}^*(\theta)) 
        \dd{((1-\alpha)F(\theta) +\Lambda(\theta) + \Psi(\theta))}
    \right]
    \mathbf{x}(\hat{\theta})\\
    &+ \int_{\hat{\theta}}^{\bar{\theta}}
    \left[
        \int_{\theta}^{\bar{\theta}}
        [-\alpha f(\tilde{\theta}) + \Lambda(\tilde{\theta})]
        \dd{\tilde{\theta}}
    \right.\\
    &\left.
        \quad\quad\quad +\int_{\theta}^{\bar{\theta}}
        \pi_{e}(\tilde{\theta}, \mathbf{e}^*(\tilde{\theta})) 
        \dd{((1-\alpha)F(\tilde{\theta}) +\Lambda(\tilde{\theta}) + \Psi(\tilde{\theta}))}
    \right]\dd{\mathbf{x}(\theta)}\\
    &-\Lambda(\bar{\theta})\pi_e(\bar{\theta}, \mathbf{e}^*(\bar{\theta}))\mathbf{x}(\bar{\theta})\\
    =& Q(\hat{\theta}) \mathbf{x}(\hat{\theta})
    -\Lambda(\bar{\theta})\pi_e(\bar{\theta}, \mathbf{e}^*(\bar{\theta}))\mathbf{x}(\bar{\theta})
    + \int_{\hat{\theta}}^{\bar{\theta}} Q(\theta) \dd{\mathbf{x}(\theta)},
\end{align*}
where I define
\begin{align*}
    Q(\theta) \coloneqq \int_{\theta}^{\bar{\theta}} \left[-\alpha f(\tilde{\theta}) + \Lambda(\tilde{\theta})\right]\dd{\tilde{\theta}}
    +\int_{\theta}^{\bar{\theta}} \pi_{e}(\tilde{\theta}, \mathbf{e}^*(\tilde{\theta}))
    \dd{((1-\alpha)F(\tilde{\theta}) +\Lambda(\tilde{\theta}) + \Psi(\tilde{\theta}))}
\end{align*}
for \(\theta\in[\hat{\theta}, \bar{\theta}]\).

\paragraph*{Proof of \(\partial{\hat{\mathcal{L}}(\mathbf{e}^*; \mathbf{e}^*\mid \Lambda, \Psi)}=0\).}

For \(\mathbf{x} = \mathbf{e}^*\),
we have
\begin{align*}
    \partial{\hat{\mathcal{L}}(\mathbf{e}^*; \mathbf{e}^*\mid \Lambda, \Psi)}
    &=Q(\hat{\theta}) \mathbf{e}^*(\hat{\theta})
    -\Lambda(\bar{\theta})\pi_e(\bar{\theta}, \mathbf{e}^*(\bar{\theta}))\mathbf{e}^*(\bar{\theta})
    + \int_{\hat{\theta}}^{\bar{\theta}} Q(\theta) \dd{\mathbf{e}^*(\theta)}
     \\
    &=Q(\hat{\theta}) \mathbf{e}^*(\hat{\theta})
    -\Lambda(\bar{\theta})\pi_e(\bar{\theta}, \mathbf{e}^*(\bar{\theta}))\mathbf{e}^*(\bar{\theta})
    + Q(\theta^*)(\bar{e}-e^*).
\end{align*}

Next, I show that
\begin{align}
    & Q(\theta^*) = 0, \label{eq:q-theta-star}\\
    \text{and}\quad
    &Q(\hat{\theta}) \mathbf{e}^*(\hat{\theta})
    =
    \Lambda(\bar{\theta})\pi_e(\bar{\theta}, \mathbf{e}^*(\bar{\theta}))\mathbf{e}^*(\bar{\theta}), \label{eq:q-theta-hat}
\end{align}
which gives
\(\partial{\hat{\mathcal{L}}(\mathbf{e}^*; \mathbf{e}^*\mid \Lambda, \Psi)} = 0\).

Since the function
\((1-\alpha)F(\theta) +\Lambda(\theta) + \Psi(\theta)\) is constant over the interval \([\theta^*, \bar{\theta}]\)
and \(\Lambda(\theta) = \alpha f(\theta)\) for \(\theta\in(\theta^*, \bar{\theta})\),
it follows that
\begin{align*}
    Q(\theta^*)
    &= \int_{\theta^*}^{\bar{\theta}}
    \left[
        -\alpha f(\theta)
        + \Lambda(\theta)
    \right]
    \dd{\theta}\\
    & \quad +\int_{\theta^*}^{\bar{\theta}} \pi_{e}(\theta, \mathbf{e}^*(\theta))
    \dd{((1-\alpha)F(\theta) +\Lambda(\theta) + \Psi(\theta))}
    \\
    &= 0,
\end{align*}
which shows equality~\eqref{eq:q-theta-star}.

To show equality~\eqref{eq:q-theta-hat},
we can write \(Q(\hat{\theta})\) as
\begin{align*}
    &Q(\hat{\theta}) = Q(\hat{\theta}) - Q(\theta^*)\\
    &= \int_{\hat{\theta}}^{\theta^*}
    [-\alpha f(\theta)
    + \Lambda(\theta)]
    \dd{\theta}
    +\int_{\hat{\theta}}^{\theta^*}
    \pi_{e}(\theta, \mathbf{e}^*(\theta))
    \dd{((1-\alpha)F(\theta)
    +\Lambda(\theta) + \Psi(\theta))}\\
    &= \int_{\hat{\theta}}^{\theta^*}
    [-\alpha f(\theta) + \Lambda(\theta)]
    \dd{\theta}
    +\int_{\hat{\theta}}^{\theta^*}
    \pi_{e}(\theta, e^*)
    \dd{((1-\alpha)F(\theta)
    +\Lambda(\theta) + \Psi(\theta))}\\
    &=\int_{\hat{\theta}}^{\theta^*}
    [-\alpha + (1-\alpha) \pi_{e}(\theta, e^*)]
    \dd{F(\theta)}
    + \int_{\hat{\theta}}^{\theta^*}
    \dv{\theta} \left(\Lambda(\theta)\pi_{e}(\theta, e^*)\right)
    \dd{\theta} 
    + \int_{\hat{\theta}}^{\theta^*}
    \pi_{e}(\theta, e^*)
    \dd{\Psi(\theta)}\\
    &=\int_{\hat{\theta}}^{\theta^*}
    [-\alpha + (1-\alpha) \pi_{e}(\theta, e^*)]
    \dd{F(\theta)}
    + \underbrace{(\Lambda(\theta^*) + \Psi(\theta^*))}_{=\alpha f(\theta^*)}\pi_{e}(\theta^*, e^*)
    - \underbrace{\Lambda(\hat{\theta})\pi_{e}(\hat{\theta}, e^*)}_{=0}\\
    &=\int_{\hat{\theta}}^{\theta^*}
    [-\alpha + (1-\alpha) \pi_{e}(\theta, e^*)]
    \dd{F(\theta)}
    + \alpha f(\theta^*)\pi_e(\theta^*, e^*),
\end{align*}
where 
\(\Lambda(\hat{\theta})\pi_{e}(\hat{\theta}, e^*) = 0\)
comes from the fact that
\(\pi_{e}(\hat{\theta}, e^*) = 0\) if \(\hat{\theta} > \underline{\theta}\)
and \(\Lambda(\hat{\theta}) = 0\) if \(\hat{\theta} = \underline{\theta}\).

If \(\theta^* < \bar{\theta}\),
which occurs when \(e^* < \underline{\mathbf{e}}(\bar{\theta})\),
then by Lemma~\ref{lem:foc} we have
both \(\Lambda(\bar{\theta}) = 0\)
and \(Q(\hat{\theta}) = 0\),
which yields equality~\eqref{eq:q-theta-hat}.
Otherwise,
if \(\theta^* = \bar{\theta}\),
we have \(\mathbf{e}^*(\hat{\theta}) = \mathbf{e}^*(\bar{\theta}) = e^*\).
Then equality~\eqref{eq:q-theta-hat} follows from
\(Q(\hat{\theta}) = \Lambda(\bar{\theta})\pi_e(\bar{\theta}, e^*)\),
which holds by our choice of \(\Lambda(\bar{\theta})\).
Since both equalities \eqref{eq:q-theta-star}~and~\eqref{eq:q-theta-hat}
are satisfied,
we have
\(\partial{\hat{\mathcal{L}}(\mathbf{e}^*; \mathbf{e}^*\mid \Lambda, \Psi)} = 0\).

\paragraph*{Proof of \(\partial{\hat{\mathcal{L}}(\mathbf{e}^*; \mathbf{x}^*\mid \Lambda, \Psi)}\leq 0\) for all \(\mathbf{x}\in\hat{\Phi}\).}

I have shown
\(Q(\hat{\theta}) = \Lambda(\bar{\theta})\pi_e(\bar{\theta}, \mathbf{e}^*(\bar{\theta}))\)
for both \(\theta^* < \bar{\theta}\)
and \(\theta^* = \bar{\theta}\).
Next, I show that
\(Q(\hat{\theta}) \geq Q(\theta)\) for all \(\theta\in[\hat{\theta}, \bar{\theta}]\),
which leads to
\begin{align*}
    \partial{\hat{\mathcal{L}}(\mathbf{e}^*; \mathbf{x}\mid \Lambda, \Psi)} 
    &= Q(\hat{\theta}) \mathbf{x}(\hat{\theta})
    -\Lambda(\bar{\theta})\pi_e(\bar{\theta}, \mathbf{e}^*(\bar{\theta}))\mathbf{x}(\bar{\theta})
    + \int_{\hat{\theta}}^{\bar{\theta}} Q(\theta) \dd{\mathbf{x}(\theta)}\\
    &= Q(\hat{\theta})
    (\mathbf{x}(\hat{\theta}) -\mathbf{x}(\bar{\theta}))
    + \int_{\hat{\theta}}^{\bar{\theta}} Q(\theta) \dd{\mathbf{x}(\theta)}\\
    &\leq 
    Q(\hat{\theta})
    (\mathbf{x}(\hat{\theta}) -\mathbf{x}(\bar{\theta}))
    + Q(\hat{\theta})\int_{\hat{\theta}}^{\bar{\theta}} \dd{\mathbf{x}(\theta)}\\
    &= 0.
\end{align*}

Since the function \((1-\alpha) F(\theta) + \Lambda(\theta) + \Psi(\theta)\)
is constant over the interval \([\theta^*, \bar{\theta}]\),
we have \(Q(\theta) = 0\)
for all \(\theta\in[\theta^*, \bar{\theta}]\) by our choice of \(\Lambda\).
Moreover, we know from Lemma~\ref{lem:foc} that \(Q(\hat{\theta}) \geq 0\),
and thus \(Q(\hat{\theta}) \geq Q(\theta)\) for all \(\theta\in[\theta^*, \bar{\theta}]\).

To show \(Q(\hat{\theta}) \geq Q(\theta)\) for all \(\theta\in(\hat{\theta}, \theta^*)\),
suppose by contradiction that
\(Q(\hat{\theta}) < Q(\theta_1)\)
for some \(\theta_1 \in(\hat{\theta}, \theta^*)\).

By the definition of \(Q\),
we know that
\(Q(\theta^*-) = Q(\theta^*)=0\),
and \(Q\) is continuously differentiable on \((\hat{\theta}, \theta^*)\),
with
\begin{align*}
    Q'(\theta) &= \alpha f(\theta) - \Lambda(\theta)\\
    &= (1-\alpha)F(\theta) + \alpha f(\theta) - (1-\alpha)F(\theta^*) - \alpha f(\theta^*).
\end{align*}
Since \(Q(\theta_1) > Q(\hat{\theta}) \geq 0 = Q(\theta^*-)\),
we must have \(Q'(\theta_2) < 0 = Q'(\theta^*-)\) for some \(\theta_2\in(\theta_1, \theta^*)\).
Also, note that
\(Q'\) is quasiconcave
due to the quasiconcavity of the function \((1-\alpha)F(\theta) + \alpha f(\theta)\).
This property of \(Q'\),
together with the inequality \(Q'(\theta_2) < Q'(\theta^*-)\),
implies that \(Q'\) must be nondecreasing on \((\hat{\theta}, \theta_2)\).
Consequently,
we have \(Q'(\theta) \leq Q'(\theta_1) \leq Q'(\theta_2)  < 0\) for all \(\theta\in(\hat{\theta}, \theta_1)\).
However,
from the definition of \(Q\)
we can observe
that \(Q(\hat{\theta}) \geq Q(\hat{\theta}+)\).\footnote{
    By the definition of \(Q\),
    we have \(Q(\hat{\theta}+) = Q(\hat{\theta})\)
    whenever \(\pi_e(\hat{\theta}, e^*) = 0\).
    The inequality might be strict
    only when \(\pi_e(\hat{\theta}, e^*) > 0\),
    which can only occur
    when \(e^* < \ehat(\underline{\theta})\).
    In such a case,
    \((1-\alpha) F(\theta) + \Lambda(\theta) + \Psi(\theta)\)
    has a nonnegative jump at \(\hat{\theta} = \underline{\theta}\).
    Therefore, \(Q\) can only have a nonpositive jump
    at \(\hat{\theta}\).
}
Then, the fact that \(Q'(\theta) < 0\) for all \(\theta\in(\hat{\theta}, \theta_1)\)
leads to \(Q(\hat{\theta})\geq Q(\theta_1)\), a contradiction.

Lastly,
I return to my earlier claim that
that the jump in \((1-\alpha) F(\theta) + \Lambda(\theta) + \Psi(\theta)\)
at \(\hat{\theta}\) is nonnegative
when \(\underline{\theta} < \hat{\theta} < \bar{\theta}\).
I continue to validate this assertion by verifying the inequality
\[
    (1-\alpha)F(\theta^*)
    +
    \alpha f(\theta^*)
    \geq
    (1-\alpha)F(\hat{\theta})
    +
    \alpha f(\hat{\theta}).
\]
From the definition of \(Q\),
we can observe that
\(Q\) is right-continuous at \(\hat{\theta}\)
if \(\hat{\theta} > \underline{\theta}\).
I have also shown that
\(Q(\hat{\theta})\geq Q(\theta)\) for all \(\theta\in[\hat{\theta}, \theta^*]\).
Therefore,
the right-hand derivative of \(Q\) at \(\hat{\theta}\) must be nonpositive,
which gives
\begin{align*}
    0 & \geq Q'(\hat{\theta}+) = (1-\alpha)F(\hat{\theta}) + \alpha f(\hat{\theta}) - (1-\alpha)F(\theta^*) - \alpha f(\theta^*).
\end{align*}
Therefore,
if \(\hat{\theta} > \underline{\theta}\),
the jump in \((1-\alpha) F(\theta) + \Lambda(\theta) + \Psi(\theta)\)
at \(\hat{\theta}\) is nonnegative.
As a result, the function \((1-\alpha) F(\theta) + \Lambda(\theta) + \Psi(\theta)\) is indeed nondecreasing on \(\Theta\).

To summarize,
I have shown that
\(\partial\hat{\mathcal{L}}(\mathbf{e}^*; \mathbf{e}^* \mid \Lambda,\Psi) = 0\)
and
\(\partial\hat{\mathcal{L}}(\mathbf{e}^*; \mathbf{x}\mid \Lambda,\Psi) \leq 0\)
for all \(\mathbf{x}\in\hat{\Phi}\).
Therefore,
according to Lemma A.2 in \textcite{Amador-Werning-Angeletos:2006},
I conclude that \(\mathbf{e}^*\) maximizes \(\hat{\mathcal{L}}\) over \(\hat{\Phi}\),
and thus also maximizes \(\mathcal{L}\) over \(\Phi\).

\paragraph*{Applying Luenberger's Sufficiency Theorem.}

Next, I apply Theorem 1 in \textcite{Amador-Bagwell:2013},
which is restated here for convenience.
\begin{theorem}[\textcite{Amador-Bagwell:2013}, Theorem 1]
    \label{thm:ab}
    Let \(f\) be a real-valued functional defined on a subset \(\Omega\)
    of a linear space \(X\).
    Let \(G\) be a mapping from \(\Omega\) into the linear space \(Z\)
    having nonempty positive cone \(P\).
    Suppose that
    \begin{enumerate}[label=(\roman*)]
        \item there exists a linear function \(T:Z\to \R\)
        such that \(T(z) \geq 0\) for all \(z\in P\),
        \item there is an element \(x_0\in\Omega\) such that 
        \[
            f(x_0) + T(G(x_0)) \leq f(x) + T(G(x))\quad \text{for all }x\in\Omega,
        \]
        \item \(-G(x_0)\in P\), and
        \item \(T(G(x_0)) = 0\).
    \end{enumerate}
    Then \(x_0\) solves
    \[
        \min f(x) \text{ subject to: } -G(x)\in P, x\in\Omega.
    \]
\end{theorem}

To apply this theorem,
I set
\begin{enumerate}[label=(\roman*)]
    \item \(x_0 \coloneqq \mathbf{e}^*\);
    \item \(X\coloneqq \{\mathbf{e}\mid \mathbf{e}:\Theta\to E\}\);
    \item \(f\) to be given by \(-\int_{\underline{\theta}}^{\bar{\theta}} [-\alpha \mathbf{e}(\theta) + (1-\alpha) \pi(\theta, \mathbf{e}(\theta))] \dd{F(\theta)}\),
    as a function of \(\mathbf{e}\in X\);
    \item \(Z\coloneqq \{(z_1, z_2, z_3)\mid z_1: \Theta\to\R, z_2:\Theta\to\R \text{ and }z_3:\Theta\to\R \text{ with }z_1, z_2, z_3 \text{ integrable}\}\);
    \item \(\Omega \coloneqq \Phi\);
    \item \(P\coloneqq\{(z_1,z_2,z_3)\mid (z_1, z_2, z_3)\in Z
    \text{ such that } z_1(\theta) \geq 0, z_2(\theta) \geq 0 \text{ and }z_3(\theta)\geq 0 \text{ for all }\theta\in\Theta\}\);
    \item \(G\) to be the mapping from \(\Phi\) to \(Z\) given by the left-hand sides of inequalities~\eqref{ineq:IC-1},~\eqref{ineq:IC-2} and~\eqref{ineq:IR};
    \item \(T\) to be the linear mapping:
    \[
        T((z_1, z_2, z_3))\coloneqq
        \int_\Theta z_1(\theta) \dd{\Lambda_1(\theta)}
        + \int_\Theta z_2(\theta) \dd{\Lambda_2(\theta)}
        + \int_\Theta z_3(\theta) \dd{\Psi(\theta)},
    \]
    where \(\Lambda_1\), \(\Lambda_2\), and \(\Psi\)
    being nondecreasing functions implies that \(T(z)\geq 0\) for \(z\in P\).
\end{enumerate}

We have that
\begin{align*}
    T(G(x_0))
    &=
    -\left[\int_\Theta
        \int_\theta^{\bar{\theta}}
        \pi_\theta(\tilde{\theta}, \mathbf{e}^*(\tilde{\theta}))
        \dd{\theta}
        + \pi(\theta, \mathbf{e}^*(\theta))
        - \pi(\bar{\theta}, \mathbf{e}^*(\bar{\theta}))
    \right]
    \dd{(\Lambda_1(\theta) - \Lambda_2(\theta))}\\
    & \quad -\int_\Theta (\pi(\theta, \mathbf{e}^*(\theta)) - \pi(\theta, \bar{e})) \dd{\Psi(\theta)} \\
    &= 0,
\end{align*}
where the last equality follows from the construction of \(\mathbf{e}^*\) and the proposed multipliers.
We have found the conditions under which the proposed \(\mathbf{e}^*\)
minimizes \(f(x) + T(G(x))\) for \(x\in\Omega\).
Given that \(T(G(x_0)) = 0\),
the conditions of Theorem~\ref{thm:ab} hold and it follows that
\(\mathbf{e}^*\) solves \(\min_{x\in\Omega} f(x)\)
subject to \(-G(x)\in P\),
which is the original problem.

\end{proof}

\begin{lemma}
    \label{lem:foc}

    Suppose \((1-\alpha)F(\theta) + \alpha f(\theta)\) is quasiconcave.
    Let \(e^*\) be the threshold of a threshold policy that solves
    Problem~\eqref{prob:threshold-policies}.
    Then, the following condition holds:
    \[
        -\int_{\hat{\theta}}^{\theta^*}
        [-\alpha + (1-\alpha) \pi_{e}(\theta, e^*)]\dd{F(\theta)}
        \begin{cases}
            =\alpha f(\theta^*)\pi_{e}(\theta^*, e^*), & \text{ if } e^* < \underline{\mathbf{e}}(\bar{\theta}),\\
            \in[0, \alpha f(\theta^*)\pi_{e}(\theta^*, e^*)], & \text{ if } e^* = \underline{\mathbf{e}}(\bar{\theta}),\\
            =0, & \text{ if } \underline{\mathbf{e}}(\bar{\theta}) < e^* \leq \bar{e}.
        \end{cases}
    \]

\end{lemma}

\begin{proof}

    The objective value of Problem~\eqref{prob:threshold-policies}
    as a function of the threshold \(e^*\)
    is given by
    \begin{align*}
        W(e^*)= &
        \int_{\underline{\theta}}^{\hat{\theta}}
        [-\alpha \ehat(\theta) + (1-\alpha) \pi(\theta, \ehat(\theta))]
        \dd{F(\theta)}\\
        &+ \int_{\hat{\theta}}^{\theta^*}
        [-\alpha e^* + (1-\alpha) \pi(\theta, e^*)]
        \dd{F(\theta)}
        + \int_{\theta^*}^{\bar{\theta}}
        [-\alpha \bar{e} + (1-\alpha) \pi(\theta, \bar{e})]
        \dd{F(\theta)}.
    \end{align*}
    Take the derivative with respect to \(e^*\), we have
    \begin{align*}
        W'(e^*) = &[-\alpha e^* + (1-\alpha)\pi(\hat{\theta}, e^*)]f(\hat{\theta})\dv{\hat{\theta}}{e^*}
        + [-\alpha e^* + (1-\alpha) \pi(\theta^*, e^*)]f(\theta^*)\dv{\theta^*}{e^*}\\
        &- [-\alpha e^* + (1-\alpha) \pi(\hat{\theta}, e^*)]f(\hat{\theta})\dv{\hat{\theta}}{e^*}
        + \int_{\hat{\theta}}^{\theta^*}
        [-\alpha + (1-\alpha) \pi_{e}(\theta, e^*)]
        \dd{F(\theta)}\\
        &- [-\alpha \bar{e} + (1-\alpha) \underbrace{\pi(\theta^*, \bar{e})}_{=\pi(\theta^*, e^*)}]f(\theta^*)\dv{\theta^*}{e^*}\\
        &=
        \alpha (\bar{e} - e^*)f(\theta^*)\dv{\theta^*}{e^*}
        + \int_{\hat{\theta}}^{\theta^*}
        [-\alpha + (1-\alpha) \pi_{e}(\theta, e^*)]
        \dd{F(\theta)}.
    \end{align*}

    \begin{enumerate}[label=(\roman*)]
        \item 
        If \(e^* < \underline{\mathbf{e}}(\bar{\theta})\),
        then \(\dv{\theta^*}{e^*} = \frac{1}{\underline{\mathbf{e}}'(\theta^*)}\).\footnote{
            Note that we must have \(e^* > \underline{\mathbf{e}}(\underline{\theta})\)
            because no disclosure is never optimal.
        }
        By the definition of \(\underline{\mathbf{e}}(\,\cdot\,)\)
        we have \(\pi(\theta, \underline{\mathbf{e}}(\theta)) = \pi(\theta, \bar{e})\).
        Taking the derivative with respect to \(\theta\) on both sides gives
        \[
            \pi_{\theta}(\theta, \underline{\mathbf{e}}(\theta))
            + \pi_{e}(\theta, \underline{\mathbf{e}}(\theta)) \underline{\mathbf{e}}'(\theta)
            = \pi_{\theta}(\theta, \bar{e}).
        \]
        Therefore,
        we have \(\underline{\mathbf{e}}'(\theta^*) = (\bar{e} - e^*)/\pi_{e}(\theta^*, e^*)\),
        and hence
        \(\dv{\theta^*}{e^*} = \frac{\pi_{e}(\theta^*, e^*)}{\bar{e} - e^*}\).
        Then,
        the first-order condition yields
        \[
            -\int_{\hat{\theta}}^{\theta^*}
            [-\alpha + (1-\alpha) \pi_{e}(\theta, e^*)]
            \dd{F(\theta)}=\alpha f(\theta^*)\pi_{e}(\theta^*, e^*).
        \]
        
        \item
        If \(e^* = \underline{\mathbf{e}}(\bar{\theta}) = \bar{e}\),
        then we have \(\hat{\theta} =\theta^* = \bar{\theta}\),
        and hence 
        \[
            -\int_{\hat{\theta}}^{\theta^*}
            [-\alpha + (1-\alpha) \pi_{e}(\theta, e^*)]\dd{F(\theta)} = 0
            \in[0, \alpha f(\theta^*)\pi_{e}(\theta^*, e^*)].
        \]
        
        \item
        If \(e^* = \underline{\mathbf{e}}(\bar{\theta}) < \bar{e}\),
        then \(W'(e^*-)\) must be nonnegative,
        which yields
        \[
            - \int_{\hat{\theta}}^{\theta^*}
            [-\alpha + (1-\alpha) \pi_{e}(\theta, e^*)]
            \dd{F(\theta)}\leq\alpha f(\theta^*)\pi_{e}(\theta^*, e^*).
        \]
        Similarly,
        \(W'(e^*+)\) must be nonpositive.
        Because \(\left.\dv{\theta^*}{e^*}\right\rvert_{e^*=\underline{\mathbf{e}}(\bar{\theta})+} = 0\),
        we have
        \[
            - \int_{\hat{\theta}}^{\theta^*}
            [-\alpha + (1-\alpha) \pi_{e}(\theta, e^*)]
            \dd{F(\theta)}\geq 0.
        \]
        
        \item
        Lastly, if \(\underline{\mathbf{e}}(\bar{\theta}) < \bar{e}\)
        and \(e^* \in (\underline{\mathbf{e}}(\bar{\theta}), \bar{e}]\),
        then we have \(\dv{\theta^*}{e^*} = 0\).
        The first-order condition then yields
        \[
            -\int_{\hat{\theta}}^{\theta^*}
            [-\alpha + (1-\alpha) \pi_{e}(\theta, e^*)]
            \dd{F(\theta)}= 0.
        \]

    \end{enumerate}
\end{proof}

\begin{lemma}
    \label{lem:peak-range}
    Suppose \((1-\alpha)F(\theta) + \alpha f(\theta)\) is quasiconcave,
    and denote the set of its maximizers by
    \(\Theta_\blacktriangle\coloneqq \arg\max_{\theta\in\Theta} \{(1-\alpha)F(\theta) + \alpha f(\theta)\}\).
    If a threshold policy
    with a threshold \(e^*\)
    solves Problem~\eqref{prob:threshold-policies},
    then we have \([\hat{\theta}, \theta^*]\cap \Theta_\blacktriangle\neq \varnothing\).

\end{lemma}

\begin{proof}

In this proof, I distinguish three cases.

\paragraph*{Case 1: \(e^* < \underline{\mathbf{e}}(\bar{\theta})\).}

Since \(e^* \leq \underline{\mathbf{e}}(\underline{\theta})\) is never optimal,
in this case we are certain that \(e^*\in(\underline{\mathbf{e}}(\underline{\theta}), \underline{\mathbf{e}}(\bar{\theta}))\).
For an arbitrary threshold policies with threshold \(e_o\in(\underline{\mathbf{e}}(\underline{\theta}), \underline{\mathbf{e}}(\bar{\theta}))\)
and the corresponding \(\hat{\theta}_o\) and \(\theta^*_o\),
the derivative of \(W\) at \(e_o\) is given by
\begin{align*}
    W'(e_o) 
    =&
    \int_{\hat{\theta}_o}^{\theta^*_o}
    [-\alpha + (1-\alpha)\pi_e(\theta, e_o)]
    \dd{F(\theta)}
    + \alpha f(\theta^*_o)\pi_e(\theta^*_o, e_o)
    \\
    =&\int_{\hat{\theta}_o}^{\theta^*_o}
    \pi_e(\theta, e_o)
    \dd{((1-\alpha)F(\theta) + \alpha f(\theta))}
    +\alpha f(\hat{\theta}_o)\pi_e(\hat{\theta}_o, e_o).
\end{align*}

At the optimal threshold \(e^*\), the first-order condition yields
\begin{align}
    W'(e^*)
    =
    \int_{\hat{\theta}}^{\theta^*}
    \pi_e(\theta, e^*)
    \dd{((1-\alpha)F(\theta) + \alpha f(\theta))}
    + \alpha f(\hat{\theta})\pi_e(\hat{\theta}, e^*)
    = 0,
    \label{eq:foc-case-1}
\end{align}
which implies
\[
    \int_{\hat{\theta}}^{\theta^*}
    \pi_e(\theta, e^*)
    \dd{((1-\alpha)F(\theta) + \alpha f(\theta))}
    = -\alpha f(\hat{\theta})\pi_e(\hat{\theta}, e^*)
    \leq 0.
\]
Now, suppose by contradiction that \([\hat{\theta}, \theta^*] \cap \Theta_\blacktriangle = \varnothing\).
Then, the quasiconcave function \((1-\alpha)F(\theta) + \alpha f(\theta)\) must be either nondecreasing or nonincreasing on the interval \([\hat{\theta}, \theta^*]\).
Moreover,
since \(\pi_e(\theta, e^*) > 0\) on \((\hat{\theta}, \theta^*)\),
the above inequality implies the function \((1-\alpha)F(\theta) + \alpha f(\theta)\)
can only be nonincreasing on \([\hat{\theta}, \theta^*]\).
It must then follow that \((1-\alpha)F(\theta) + \alpha f(\theta)\)
is constant on \([\hat{\theta}, \theta^*]\).
To see this,
suppose by contradiction that
this function contains a strictly decreasing part within \([\hat{\theta}, \theta^*]\).
Then the integral above has to be negative,
which implies \(\pi_e(\hat{\theta}, e^*)\) is positive.
Such a scenario can only occur if \(e^* < \ehat(\underline{\theta})\),
in which case \(\hat{\theta} = \underline{\theta}\).
However,
since \((1-\alpha)F(\theta) + \alpha f(\theta)\)
is quasiconcave and nonincreasing on \([\hat{\theta}, \theta^*]\),
if \(\hat{\theta} = \underline{\theta}\),
then \(\hat{\theta}\) has to be a maximizer,
contradicting the hypothesis that \([\hat{\theta}, \theta^*] \cap \Theta_\blacktriangle = \varnothing\).

Therefore, if \([\hat{\theta}, \theta^*] \cap \Theta_\blacktriangle = \varnothing\),
then \((1-\alpha)F(\theta) + \alpha f(\theta)\) has to be constant on \([\hat{\theta}, \theta^*]\).
It then follows that the integral above is zero.
However,
since the maximizers are outside the interval \([\hat{\theta}, \theta^*]\)
and \((1-\alpha)F(\theta) + \alpha f(\theta)\) is quasiconcave on \(\Theta\),
we can increase the threshold from \(e^*\) until the derivative becomes positive if those maximizers are on the right,
or decrease the threshold until the derivative becomes negative if they are on the left.\footnote{
    Since both the derivative
    and the integral in \eqref{eq:foc-case-1} are zero,
    the second term in \eqref{eq:foc-case-1},
    \(\alpha f(\hat{\theta}) \pi_e(\hat{\theta}, e^*)\),
    must be zero as well,
    either because \(\hat{\theta} = \underline{\theta}\) and \(\alpha f(\underline{\theta}) = 0\),
    or \(\hat{\theta} > \underline{\theta}\) and hence \(\pi_e(\hat{\theta}, e^*) = 0\).
    In either case, the second term in the derivative
    remains at zero as we increase the threshold from \(e^*\)
    when the maximizers lie on the right of the interval \([\hat{\theta}, \theta^*]\).
    If, on the other hand, the maximizers are on the left of \([\hat{\theta}, \theta^*]\),
    then we must have \(\hat{\theta} > \underline{\theta}\),
    and thus the second term stays at zero
    throughout this process.
}
Either of these two cases contradicts the optimality of \(e^*\).

\paragraph*{Case 2: \(\underline{\mathbf{e}}(\bar{\theta})\leq e^* < \ehat(\bar{\theta})\).}

If \(\hat{\theta} = \underline{\theta}\),
then we have \([\hat{\theta}, \theta^*] = [\underline{\theta},\bar{\theta}]\),
and hence the statement holds trivially.

Suppose \(\hat{\theta} > \underline{\theta}\),
which implies \(\pi_e(\hat{\theta}, e^*) = 0\).
The first-order condition yields
\begin{align*}
    W'(e^*) &= \int_{\hat{\theta}}^{\theta^*}
    [-\alpha + (1-\alpha) \pi_{e}(\theta, e^*)]\dd{F(\theta)}\\
    &= \int_{\hat{\theta}}^{\theta^*}
    \pi_e(\theta, e^*)
    \dd{((1-\alpha)F(\theta) + \alpha f(\theta))}
    - \alpha f(\theta^*) \pi_e(\theta^*, e^*)\\
    &= 0,
\end{align*}
which implies
\begin{align*}
    \int_{\hat{\theta}}^{\theta^*}
    \pi_e(\theta, e^*)
    \dd{((1-\alpha)F(\theta) + \alpha f(\theta))}
    = \alpha f(\theta^*) \pi_e(\theta^*, e^*)
    \geq 0.
\end{align*}

Suppose by contradiction that \([\hat{\theta}, \theta^*] \cap \Theta_\blacktriangle = \varnothing\).
Then
\((1-\alpha)F(\theta) + \alpha f(\theta)\) is nonincreasing 
on \([\hat{\theta}, \theta^*]\),
because it is quasiconcave and all its maximizers lie to the left of
the interval \([\hat{\theta}, \theta^*] = [\hat{\theta}, \bar{\theta}]\).
As \(\pi_e(\theta, e^*) > 0\) on \((\hat{\theta}, \theta^*)\),
the above inequality further implies \((1-\alpha)F(\theta) + \alpha f(\theta)\) is constant on \([\hat{\theta}, \theta^*]\),
and consequently the inequality holds with equality.
Thus, both the integral and the second term in the derivative are zero.
Similar to the previous case,
decreasing the threshold \(e^*\) eventually leads to negative derivatives,
either because the integral becomes negative,
or because the second term in the derivative becomes positive.\footnote{
    We might need to lower the threshold blow
    \(\underline{\mathbf{e}}(\bar{\theta})\),
    which brings us back to Case 1.
    The second term in the derivative
    then becomes \(\alpha f(\hat{\theta}_o)\pi_e(\hat{\theta}_o, e_o)\),
    which stays at zero throughout the process.
}
This observation contradicts the optimality of the threshold \(e^*\).

\paragraph*{Case 3: \(\ehat(\bar{\theta}) \leq e^* \leq \bar{e}\).}
In this case,
I assume without loss of generality
that \(e^* = \ehat(\bar{\theta})\).

Suppose by contradiction that \([\hat{\theta}, \theta^*] \cap \Theta_\blacktriangle = \varnothing\).
In this case, \([\hat{\theta}, \theta^*] = \{\bar{\theta}\}\),
and thus any element in \(\Theta_\blacktriangle\) must lie strictly below \(\bar{\theta}\).
Hence, there exists an interval \((\bar{\theta}-\delta ,\bar{\theta}]\) for some \(\delta > 0\),
such that \((\bar{\theta}-\delta ,\bar{\theta}]\cap \Theta_\blacktriangle = \varnothing\),
and
the function \((1-\alpha)F(\theta) + \alpha f(\theta)\) is nonincreasing
on \((\bar{\theta}-\delta ,\bar{\theta}]\)
as a consequence of its quasiconcavity.
Since \(e^* = \ehat(\bar{\theta})\) is optimal,
we have
\[
    W'(e_o)
    = \int_{\hat{\theta}_o}^{\theta^*_o}
    \pi_e(\theta, e_o)
    \dd{((1-\alpha)F(\theta) + \alpha f(\theta))}
    - \alpha f(\theta^*_o)\pi_e(\theta^*_o, e_o)
    \ind_{\{\theta^*_o = \bar{\theta}\}}
    \geq 0
\]
for all \(e_o\) less than but sufficiently close to \(e^*\).
However,
since \((1-\alpha)F(\theta) + \alpha f(\theta)\) is nonincreasing
on \((\bar{\theta} - \delta, \bar{\theta}]\),
for all \(e_o\) sufficiently close to \(e^* = \ehat(\bar{\theta})\)
with \(\hat{\theta}_o > \bar{\theta} - \delta\),
the integral above must be nonpositive,
and thus the inequality must hold with equality.
Therefore,
the derivatives \(W'(e_o) = 0\) for all \(e_o\) sufficiently close to \(e^* = \ehat(\bar{\theta})\),
which implies 
some threshold policy,
with a threshold \(e_o < \ehat(\bar{\theta})\)
and \([\hat{\theta}_o, \theta^*_o]\subset (\bar{\theta}-\delta,\bar{\theta}]\),
is also optimal within the class of threshold policies.
Since 
\((\bar{\theta} - \delta, \bar{\theta}] \cap \Theta_\blacktriangle = \varnothing\),
we also have \([\hat{\theta}_o, \theta^*_o] \cap \Theta_\blacktriangle = \varnothing\).
However,
this is a contradiction,
because it has been shown
(in Case 1 if \(\underline{\mathbf{e}}(\bar{\theta}) = \ehat(\bar{\theta}) = \bar{e}\), and in Case 2 if \(\underline{\mathbf{e}}(\bar{\theta}) < \ehat(\bar{\theta}) < \bar{e}\))
that
if \(e_o\) is an optimal threshold
with \(e_o < \ehat(\bar{\theta})\),
we must have \([\hat{\theta}_o, \theta^*_o]\cap \Theta_\blacktriangle\neq \varnothing\).

\end{proof}

\section{Proof of Theorem~\ref{thm:frontier-char}}
\label{sec:proof-thm-frontier-char}
\begin{proof}[Proof of Theorem~\ref{thm:frontier-char}.]
    Let \(d_{e^*}\in \mathcal{D}^*\) denote a threshold policy with a generic threshold \(e^*\in E\).
    Let \(d_{e_1}\) be a solution to Problem~\eqref{prob:threshold-policies} for \(\alpha = 1\).
    Note that the set \(\{(\Gamma(d_{e^*}), \Pi(d_{e^*}))\in\mathbb{R}^2 \mid d_{e^*}\in\mathcal{D}^*, e^*\in[e_1, \bar{e}]\}\)
    can be represented by a continuous curve in \(\mathbb{R}^2\)
    parametrized by the thresholds \(e^*\in[e_1, \bar{e}]\).\footnote{
        This set may contain only one point.
    }
    We know from Proposition~\ref{prop:full-disclosure-profit-maximization}
    that \(d_{\bar{e}}\) solves Problem~\eqref{prob:all-policies} for \(\alpha = 0\).
    Lemma~\ref{lem:global-sufficiency} implies
    that \(d_{e_1}\) solves Problem~\eqref{prob:all-policies} for \(\alpha = 1\).
    Therefore,
    the emission-profit pair induced by any disclosure policy in \(\mathcal{D}\)
    must lie in the region 
    \(\left[\Gamma(d_{e_1}), +\infty\right)\times\left(-\infty,  \Pi(d_{\bar{e}})\right]\).

    Suppose by contradiction that
    for some efficient policy \(d_P\in\mathcal{D}\),
    its emission-profit pair \((\Gamma(d_P), \Pi(d_P))\)
    cannot be induced by any threshold policy.
    Then by the continuity of the above-mentioned parametric curve,
    there must exist a threshold policy \(d_{e^*}\) with a threshold \(e^*\in[e_1,\bar{e}]\)
    that is strictly dominated by \(d_P\) in both dimensions,
    i.e.,
    \(\Gamma(d_{e^*}) > \Gamma(d_P)\) and \(\Pi(d_{e^*}) < \Gamma(d_P)\).
    This observation implies that \(d_{e^*}\) cannot be a solution to Problem~\eqref{prob:all-policies} for any \(\alpha\in[0,1]\),
    which contradicts Lemma \ref{lem:global-sufficiency} and~\ref{lem:threshold-policy-optimal-for-some-alpha}.

    The uniqueness of the threshold 
    follows from Lemma~\ref{lem:obj-strictly-quasiconcave},
    as the strict log-concavity of \(f\) implies
    the strict quasiconcavity of \((1-\alpha) F(\theta) + \alpha f(\theta)\) for any \(\alpha\in[0,1]\).
\end{proof}
\begin{lemma}
    \label{lem:threshold-policy-optimal-for-some-alpha}
    Suppose the function \((1-\alpha)F(\theta) + \alpha f(\theta)\) is quasiconcave for all \(\alpha\in[0,1]\).
    Let \(e_1\) be the threshold of any threshold policy that solves Problem~\eqref{prob:threshold-policies} for \(\alpha = 1\).
    For any \(e^*\in[e_1, \bar{e}]\),
    there exists an \(\alpha\in[0,1]\)
    such that a threshold policy with threshold \(e^*\) solves Problem~\eqref{prob:threshold-policies}.
\end{lemma}

\begin{proof}
    The claim is trivial for \(e^* = e_1\) by the definition of \(e_1\).
    Additionally,
    Proposition~\ref{prop:full-disclosure-profit-maximization}
    asserts that 
    any threshold policy with threshold \(e^*\in[\hat{\mathbf{e}}(\bar{\theta}), \bar{e}]\)
    solves Problem~\eqref{prob:threshold-policies} for \(\alpha = 0\).
    It remains to prove the claim for \(e^*\in(e_1, \hat{\mathbf{e}}(\bar{\theta}))\).

    I will first prove this result
    under the assumption that the function \((1-\alpha)F(\theta) + \alpha f(\theta)\)
    is strictly quasiconcave,
    and then extend the result to the case
    where it is only required to be quasiconcave,
    as stated in the Lemma.

    Suppose \((1-\alpha)F(\theta) + \alpha f(\theta)\) is strictly quasiconcave on \(\Theta\).
    Let \(e^*\in(e_1, \hat{\mathbf{e}}(\bar{\theta}))\).
    For each \(\alpha\),
    the right-hand derivative of \(W\) at \(e^*\) is given by
    \[
        W'(e^*+\,;\, \alpha) = \int_{\hat{\theta}}^{\theta^*}
        [-\alpha + (1-\alpha) \pi_{e}(\theta, e^*)]\dd{F(\theta)} 
        + \alpha f(\theta^*)\pi_{e}(\theta^*, e^*)\ind_{\{e^* < \underline{\mathbf{e}}(\bar{\theta})\}}.
    \]
    From this expression,
    it is obvious that \(W'(e^*+\,;\, \alpha)\geq 0\) for \(\alpha = 0\).
    For \(\alpha = 1\),
    we have \(W'(e^*+\,;\, \alpha) \leq 0\),
    because 
    \(e^*\) lies to the right of the global maximizer \(e_1\),
    and
    \(W(\,\cdot\,;\, \alpha)\) is strictly quasiconcave
    on \((\underline{\mathbf{e}}(\underline{\theta}), \ehat(\bar{\theta}))\)
    by Lemma~\ref{lem:obj-strictly-quasiconcave}.
    Therefore,
    by the continuity of \(W'(e^*+\,;\,\alpha)\) as a function of \(\alpha\),
    we have \(W'(e^*+\,;\, \alpha) = 0\) for some \(\alpha \in[0,1]\),
    which,
    by Lemma~\ref{lem:obj-strictly-quasiconcave},
    implies that \(e^*\) solves Problem~\eqref{prob:threshold-policies}.

    Next,
    instead of assuming \((1-\alpha)F(\theta) + \alpha f(\theta)\)
    is strictly quasiconcave,
    suppose it is quasiconcave on \(\Theta\).

    Let \(\{f_n\}_{n\in\mathbb{N}}\) be a sequence of probability density functions that converge pointwise to \(f\) on \(\Theta\),
    with each \(f_n\) continuously differentiable on \(\interior\Theta\)
    and \((1-\alpha)F_n(\theta) + \alpha f_n(\theta)\) strictly quasiconcave on \(\Theta\).\footnote{
        For example, one can define
        \(f_n(\theta)\coloneqq c_n f(\theta)(1-(\theta- \theta_\blacktriangle)^2/(n+N))\) for some normalizing factor \(c_n > 0\) and a sufficiently large integer \(N\),
        where \(\theta_\blacktriangle\) maximizes \((1-\alpha)F(\theta) + \alpha f(\theta)\) over \(\Theta\).
    }
    For each \(f_n\) and \(\alpha\in[0,1]\),
    denote by
    \[
        W_n(e\,;\,f_n, \alpha)\coloneqq\int_\Theta (-\alpha \mathbf{e}(\theta) + (1-\alpha) \pi(\theta, \mathbf{e}(\theta))) f_n(\theta)\dd{\theta}
    \]
    the objective value of Problem~\eqref{prob:threshold-policies}
    evaluated at a threshold policy with a threshold \(e\in(\underline{\mathbf{e}}(\underline{\theta}), \ehat(\bar{\theta}))\),
    under the type distribution \(f_n\).
    Let \(e^*\in(e_1, \hat{\mathbf{e}}(\bar{\theta}))\).
    Since \((1-\alpha)F_n(\theta) + \alpha f_n(\theta)\)
    is strictly quasiconcave,
    based on the first part of this proof,
    we know there exists some \(\alpha_n\in[0, 1]\)
    such that \(W_n(\,\cdot\,;\, f_n, \alpha_n)\) attains its maximum at \(e^*\).

    Note that the sequence \(\{\alpha_n\}_{n\in\mathbb{N}}\) converges to some \(\alpha^*\in[0,1]\).
    To see this,
    observe that for each fixed \(e^*\),
    the expression for \(W'(e^*+\,;\, \alpha)\) above
    is linear in \(\alpha\),
    and thus
    each \(W_n'(e^*+\,;\, f_n, \alpha)\)
    can be written as \(k_n \alpha + b_n\)
    for some constants \(k_n\) and \(b_n\).
    Moreover, it is clear that
    \(
        b_n = \int_{\hat{\theta}}^{\theta^*} \pi_e(\theta, e^*) \dd{F_n(\theta)}
    \)
    is positive.
    Then,
    we must have \(k_n < 0\),
    because we know \(k_n + b_n \leq 0 \leq b_n\),
    as we have argued in the first part of the proof.
    Thus,
    by Lemma~\ref{lem:obj-strictly-quasiconcave},
    each \(\alpha_n\) is
    the unique zero of the function \(k_n \alpha + b_n\)
    and lies within \([0, 1]\).
    As both sequences \(\{k_n\}_{n\in\mathbb{N}}\) and \(\{b_n\}_{n\in\mathbb{N}}\) are convergent
    by the dominated convergence theorem,
    we may conclude that \(\alpha^*\coloneqq\lim\alpha_n\) exists and lies within \([0,1]\).

    Hence,
    the sequence of functions
    \(\{(-\alpha_n \mathbf{e}(\theta) + (1-\alpha_n) \pi(\theta, \mathbf{e}(\theta))) f_n(\theta)\}_{n\in\mathbb{N}}\)
    converges pointwise to
    \((-\alpha \mathbf{e}(\theta) + (1-\alpha) \pi(\theta, \mathbf{e}(\theta))) f(\theta)\).
    As a result, the dominated convergence theorem implies that
    \(W_n(\,\cdot\,;\, f_n, \alpha_n)\) converges pointwise to \(W(\,\cdot\,;\, f, \alpha^*)\).
    Therefore,
    for any \(e\in E\),
    we can conclude \(W(e^*\,;\, f, \alpha^*)\geq W(e\,;\, f, \alpha^*)\),
    because \(W_n(e^*\,;\, f_n, \alpha_n) \geq W_n(e\,;\,f_n, \alpha_n)\) for each \(n\).
    This completes the proof.

\end{proof}
\begin{lemma}
    \label{lem:obj-strictly-quasiconcave}
    Suppose \(f\) is continuous differentiable on \(\interior\Theta\),
    and the function \((1-\alpha) F(\theta) + \alpha f(\theta)\) is strictly quasiconcave on \(\Theta\).
    Then, \(W\) is strictly quasiconcave on \((\underline{\mathbf{e}}(\underline{\theta}), \ehat(\bar{\theta}))\).
    For any \(e^*\in(\underline{\mathbf{e}}(\underline{\theta}), \ehat(\bar{\theta}))\),
    if either \(W'(e^*-) = 0\) or \(W'(e^*+) = 0\) holds,
    then \(e^*\) is the unique maximizer of \(W\) over \(E\).

\end{lemma}

\begin{proof}
    Note that \(W'(e^*-)\neq W'(e^*+)\)
    can only occur at \(e^* = \underline{\mathbf{e}}(\bar{\theta})\)
    when \(\underline{\mathbf{e}}(\bar{\theta}) < \bar{e}\).
    However,
    as we will see next,
    it is always the case that
    \(W'(\underline{\mathbf{e}}(\bar{\theta})-) \geq W'(\underline{\mathbf{e}}(\bar{\theta})+)\).
    This implies that,
    to prove both statements in the Lemma,
    it suffices to show that
    \(W''(e^*-) < 0\) whenever \(W'(e^*-)=0\),
    and 
    \(W''(e^*+) < 0\) whenever \(W'(e^*+)=0\).
    These conditions imply that any local extremum of \(W\) in \((\underline{\mathbf{e}}(\underline{\theta}), \ehat(\bar{\theta}))\) is a strict local maximum,
    which implies \(W\) is strictly quasiconcave in this region.
    Therefore,
    if a local maximum exists in \((\underline{\mathbf{e}}(\underline{\theta}), \ehat(\bar{\theta}))\),
    it must be the unique global maximum on \([\underline{\mathbf{e}}(\underline{\theta}), \ehat(\bar{\theta})]\),
    and,
    because \(W\) is constant outside of this region,
    on \(E\) as well.

    In what follows, I show that
    \(W''(e^*-) < 0\) whenever \(W'(e^*-)=0\),
    as well as
    \(W''(e^*+) < 0\) whenever \(W'(e^*+)=0\).
    Let \(h(\theta)\coloneqq (1-\alpha)F(\theta) + \alpha f(\theta)\),
    and consider three cases.

    \paragraph*{Case 1: \(e^* \in(\underline{\mathbf{e}}(\underline{\theta}), \underline{\mathbf{e}}(\bar{\theta}))\).}

    The first derivative of \(W\) with respect to \(e^*\) is given by
    \begin{align*}
        W'(e^*) 
        =& \int_{\hat{\theta}}^{\theta^*}
        \pi_e(\theta, e^*)
        \dd{h(\theta)}
        + \alpha f(\hat{\theta}) \pi_e(\hat{\theta}, e^*)\ind_{\{e^* \leq \ehat(\underline{\theta})\}}.
    \end{align*}
    The left-hand side second derivative is given by
    \begin{align*}
        W''(e^*-)
        =& \int_{\hat{\theta}}^{\theta^*}
        \pi_{ee}(\theta, e^*)
        \dd{h(\theta)}\\
        &\quad + \pi_e(\theta^*, e^*)h(\theta^*)\theta^{*\prime}(e^*)
        - \pi_e(\hat{\theta}, e^*)h(\hat{\theta})\hat{\theta}'(e^*-)\\
        &\quad +\alpha (f(\hat{\theta}) + f'(\hat{\theta})\pi_e(\hat{\theta}, e^*))
        \hat{\theta}'(e^*-)\ind_{\{e^* \leq \ehat(\underline{\theta})\}}\\
        &\quad
        +
        \alpha f(\hat{\theta})\pi_{ee}(\hat{\theta}, e^*)\ind_{\{e^* \leq \ehat(\underline{\theta})\}}\\
        &=\pi_0''(e^*)\left(
            h(\theta^*) - h(\hat{\theta}) + \alpha f(\hat{\theta})\ind_{\{e^* \leq \ehat(\underline{\theta})\}}
        \right)
        + \pi_e(\theta^*, e^*)h'(\theta^*)\theta^{*\prime}(e^*),
    \end{align*}
    where we use that \(\pi_e(\hat{\theta}, e^*)\hat{\theta}'(e^*-) = 0\)
    for all \(e^*\in(\underline{\mathbf{e}}(\underline{\theta}), \hat{\mathbf{e}}(\bar{\theta}))\),
    and that \(\hat{\theta}'(e^*-) = 0\) for \(e^* \leq \ehat(\underline{\theta})\).
    Since \(h\) is strictly quasiconcave,
    we must have \(h'(\theta^*) \leq 0\),
    because otherwise the expression for \(W'(e^*)\) above would be positive.
    Hence,
    the last term of \(W''(e^*-)\)
    is nonpositive,
    as both \(\pi_e(\theta^*, e^*)\)
    and \(\theta^{*\prime}(e^*)\) are positive.
    For \(e^* \leq \ehat(\underline{\theta})\),
    clearly \(h(\theta^*) - h(\hat{\theta}) + \alpha f(\hat{\theta})
    = (1-\alpha)(F(\theta^*) - F(\hat{\theta})) + \alpha f(\theta^*)\)
    is positive,
    and hence \(W''(e^* -) < 0\).
    For \(e^* > \ehat(\underline{\theta})\),
    given that \(W'(e^*) = 0\),
    Lemma~\ref{lem:peak-boundaries-relation} implies that
    \(h(\theta^*) - h(\hat{\theta}) > 0\),
    which also gives \(W''(e^* -) < 0\).

    The argument for
    \[
        0 > W''(e^*+) = \pi_0''(e^*)\left(
            h(\theta^*) - h(\hat{\theta}) + \alpha f(\hat{\theta})\ind_{\{e^* < \ehat(\underline{\theta})\}}
        \right)
        + \pi_e(\theta^*, e^*)h'(\theta^*)\theta^{*\prime}(e^*)
    \]
    is analogous and thus omitted here.

    \paragraph*{Case 2: \(e^* \in (\underline{\mathbf{e}}(\bar{\theta}), \ehat(\bar{\theta}))\).}

    The first derivatives of \(W\) with respect to \(e^*\) is given by
    \begin{align*}
        W'(e^*) 
        =& \int_{\hat{\theta}}^{\theta^*}
        \pi_e(\theta, e^*)
        \dd{h(\theta)}
        - \alpha f(\theta^*) \pi_e(\theta^*, e^*)
        + \alpha f(\hat{\theta}) \pi_e(\hat{\theta}, e^*)\ind_{\{e^* \leq \ehat(\underline{\theta})\}}.
    \end{align*}
    The right-hand side second derivative is given by
    \begin{align*}
        W''(e^*+)
        =& \int_{\hat{\theta}}^{\theta^*}
        \pi_{ee}(\theta, e^*)
        \dd{h(\theta)}\\
        &\quad + \pi_e(\theta^*, e^*)h(\theta^*)\theta^{*\prime}(e^*)
        - \pi_e(\hat{\theta}, e^*)h(\hat{\theta})\hat{\theta}'(e^*+)\\
        &\quad - \alpha (f(\theta^*) + f'(\theta^*)\pi_e(\theta^*, e^*))\theta^{*\prime}(e^*) \\
        &\quad -\alpha f(\theta^*)\pi_{ee}(\theta^*, e^*)\\
        &\quad +\alpha (f(\hat{\theta}) + f'(\hat{\theta})\pi_e(\hat{\theta}, e^*))
        \hat{\theta}'(e^*+)\ind_{\{e^* < \ehat(\underline{\theta})\}}\\
        &\quad
        +
        \alpha f(\hat{\theta})\pi_{ee}(\hat{\theta}, e^*)\ind_{\{e^* < \ehat(\underline{\theta})\}}\\
        &=\pi_0''(e^*)\left(
            h(\theta^*) - h(\hat{\theta}) -\alpha f(\theta^*) + \alpha f(\hat{\theta})\ind_{\{e^* < \ehat(\underline{\theta})\}}
        \right),
    \end{align*}
    where we use that \(\theta^{*\prime}(e^*) = 0\) for \(e^* > \underline{\mathbf{e}}(\bar{\theta})\),
    that \(\pi_e(\hat{\theta}, e^*)\hat{\theta}'(e^*+) = 0\)
    for all \(e^*\in(\underline{\mathbf{e}}(\underline{\theta}), \hat{\mathbf{e}}(\bar{\theta}))\),
    and that \(\hat{\theta}'(e^*+) = 0\) for \(e^* < \ehat(\underline{\theta})\).
    For \(e^* < \ehat(\underline{\theta})\),
    clearly \(h(\theta^*) - h(\hat{\theta}) -\alpha f(\theta^*) + \alpha f(\hat{\theta}) = (1-\alpha)(F(\theta^*) - F(\hat{\theta}))\) is positive,
    and thus \(W''(e^*+) < 0\).\footnote{
        Note that we must have \(\alpha < 1\),
        since
        we would have \(W'(e^*)=\int_{\hat{\theta}}^{\theta^*}
        [-\alpha + (1-\alpha) \pi_{e}(\theta, e^*)]\dd{F(\theta)} > 0\)
        if \(\alpha = 1\),
        which contradicts the premise that \(W'(e^*) = 0\).
    }
    For \(e^* \geq \ehat(\underline{\theta})\),
    given that \(W'(e^*)=0\),
    Lemma~\ref{lem:peak-boundaries-relation}
    implies
    \(h(\theta^*) - h(\hat{\theta}) -\alpha f(\theta^*) > 0\),
    which also gives \(W''(e^*+) < 0\).

    The argument for
    \[
        0 > W''(e^*-) = \pi_0''(e^*)\left(
            h(\theta^*) - h(\hat{\theta}) -\alpha f(\theta^*) + \alpha f(\hat{\theta})\ind_{\{e^* \leq \ehat(\underline{\theta})\}}
        \right)
    \]
    is analogous and thus omitted here.

    \paragraph*{Case 3: \(e^* = \underline{\mathbf{e}}(\bar{\theta}) < \bar{e}\).}
    If \(W'(e^*-) = 0\),
    then \(W''(e^*-) < 0\) follows from the same argument in Case 1.
    Similarly, if \(W'(e^*+) = 0\),
    then \(W''(e^*+) < 0\) follows from the same argument in Case 2.

\end{proof}
\begin{lemma}
    \label{lem:peak-boundaries-relation}
    Suppose \(h(\theta)\coloneqq (1-\alpha)F(\theta) + \alpha f(\theta)\) is quasiconcave on \(\Theta\).
    Consider a threshold policy with threshold \(e^*\in (\ehat(\underline{\theta}),\ehat(\bar{\theta}))\).
    \begin{itemize}
        \item 
        If \(W'(e^*-) = 0\), then we have
        \(h(\theta^*) \geq h(\hat{\theta}) + \alpha f(\theta^*) \ind_{\{e^* > \underline{\mathbf{e}}(\bar{\theta})\}}\);
        \item 
        If \(W'(e^*+) = 0\), then we have
        \(h(\theta^*) \geq h(\hat{\theta}) + \alpha f(\theta^*) \ind_{\{e^* \geq \underline{\mathbf{e}}(\bar{\theta})\}}\).
    \end{itemize}
    These inequalities are strict if \(h\) is strictly quasiconcave on \(\Theta\).
\end{lemma}

\begin{proof}
    Let \(e^* \in (\ehat(\underline{\theta}), \ehat(\bar{\theta}))\).
    Suppose \(W'(e^*-) = 0\), which yields
    \begin{align*}
        0
        =
        W'(e^*-)
        =
        \int_{\hat{\theta}}^{\theta^*}
        \pi_e(\theta, e^*)
        \dd{h(\theta)}
        - \alpha f(\theta^*)\pi_e(\theta^*, e^*)\ind_{\{e^* > \underline{\mathbf{e}}(\bar{\theta})\}}.
    \end{align*}
    Substituting \(\pi_e(\theta, e^*) = \theta - \hat{\theta}\),
    we obtain
    \begin{align*}
        \int_{\hat{\theta}}^{\theta^*}
        (\theta - \hat{\theta})
        \dd{h(\theta)}
        = \alpha f(\theta^*)(\theta^* - \hat{\theta})\ind_{\{e^* > \underline{\mathbf{e}}(\bar{\theta})\}}.
    \end{align*}
    Integration by parts gives
    \begin{align*}
        h(\theta^*)
        =&
        \frac{1}{\theta^* - \hat{\theta}}\int_{\hat{\theta}}^{\theta^*}
        h(\theta)
        \dd{\theta}
        +
        \alpha f(\theta^*)\ind_{\{e^* > \underline{\mathbf{e}}(\bar{\theta})\}}.
    \end{align*}

    Suppose \(h\) is quasiconcave.
    Then, we have
    \begin{align}
        h(\theta^*) 
        -
        \alpha f(\theta^*)\ind_{\{e^* > \underline{\mathbf{e}}(\bar{\theta})\}}
        &=
        \frac{1}{\theta^* - \hat{\theta}}\int_{\hat{\theta}}^{\theta^*}
        h(\theta)
        \dd{\theta}\nonumber \\
        &\geq
        \min_{\theta\in[\hat{\theta}, \theta^*]} h(\theta) \label{eq:h-min}\\
        &=
        \min\{h(\hat{\theta}), h(\theta^*)\}\nonumber \\
        &= h(\hat{\theta}), \nonumber
    \end{align}
    where the second-to-last equality
    follows from the quasiconcavity of \(h\).
    The last equality comes from the observation that 
    \(h(\hat{\theta}) \leq h(\theta^*)\),
    because otherwise
    we would have
    \[
        h(\hat{\theta})
        >
        h(\theta^*)
        \geq
        \frac{1}{\theta^* - \hat{\theta}}\int_{\hat{\theta}}^{\theta^*}
        h(\theta)
        \dd{\theta},
    \]
    which cannot hold given that \(h\) is continuous and quasiconcave on \([\hat{\theta}, \theta^*]\).

    If \(h\) is strictly quasiconcave,
    then \(h\) cannot be constant on \([\hat{\theta}, \theta^*]\),
    and thus inequality~\eqref{eq:h-min} becomes strict.

    With
    \[
        W'(e^*+)
        =
        \int_{\hat{\theta}}^{\theta^*}
        \pi_e(\theta, e^*)
        \dd{h(\theta)}
        - \alpha f(\theta^*)\pi_e(\theta^*, e^*)\ind_{\{e^* \geq \underline{\mathbf{e}}(\bar{\theta})\}},
    \]
    the proof for the case when \(W'(e^*+) = 0\) is analogous and thus omitted.

\end{proof}

\section{Proof of Proposition~\ref{prop:full-disclosure}}
\begin{proof}
    In this proof, I assume \(a < 0\).
    The case for \(a > 0\) can be shown analogously.

    First, note that firms with type \(\theta\in[\theta_\blacktriangle, \bar{\theta}]\)
    must choose \(\bar{e}\) in any equilibrium,
    regardless of the disclosure policy.
    Consequently, these types can be ignored without loss of generality,
    allowing us to assume 
    that \(\theta_\blacktriangle = \bar{\theta}\).
    In such a case,
    we have
    \(\underline{\mathbf{e}}(\bar{\theta}) = \ehat(\bar{\theta})=\bar{e}\),
    with \(f\) being nondecreasing on \(\Theta\).

    To prove the proposition,
    it suffices to show that
    full disclosure solves Problem~\eqref{prob:all-policies}
    for \(\alpha = 1\).
    This would imply that,
    in addition to maximizing the expected profit across all policies,
    full disclosure also induces the lowest equilibrium emission,
    thereby reducing the Pareto frontier to a single point.
    Any policy in \(\mathcal{D}\)
    other than full disclosure
    would induce a strictly higher expected emission
    due to the strict concavity of \(\pi(\theta, \,\cdot\,)\),
    and thus establishing the proposition.\footnote{
        A disclosure policy in \(\mathcal{D}\)
        that yields the same expected profit
        as full disclosure obviously has to implement an emission scheme \(\mathbf{e}\)
        that equals to \(\ehat\) almost everywhere.
        Then, \(\mathbf{e}\) must be equal to \(\ehat\) on \(\Theta\) everywhere.
        Because otherwise
        \eqref{eq:IC} would be violated:
        there would exist a firm with type \(\theta\)
        that could obtain a profit arbitrarily close to 
        \(\pi(\theta,\ehat(\theta)) > \pi(\theta,\mathbf{e}(\theta))\)
        by reporting some types in the neighborhood of \(\theta\).
    }
    Since \(f\) is nondecreasing,
    it is quasiconcave.
    Therefore, by Lemma~\ref{lem:global-sufficiency},
    it remains to show that
    full disclosure solves Problem~\eqref{prob:threshold-policies}
    for \(\alpha = 1\).

    Given \(\alpha = 1\),
    the derivative of \(W\)
    with respect to a threshold \(e^*\in(\underline{\mathbf{e}}(\bar{\theta}), \bar{e})\) is given by
    \[
        W'(e^*) 
        = \int_{\hat{\theta}}^{\theta^*}
        \pi_e(\theta, e^*)
        \dd{f(\theta)}
        +  f(\hat{\theta}) \pi_e(\hat{\theta}, e^*)\ind_{\{e^* \leq \ehat(\underline{\theta})\}}.
    \]
    Since \(f\) is nondecreasing on \(\Theta\),
    we have \(W'(e^*) \geq 0\).
    This implies that
    full disclosure,
    which corresponds to a threshold policy with threshold \(\bar{e}\),
    solves Problem~\eqref{prob:threshold-policies},
    as any solutions to this problem
    must have a threshold within the interval \([\underline{\mathbf{e}}(\underline{\theta}), \bar{e}]\).
    This completes the proof.

\end{proof}

\section{Proof of Lemma~\ref{lem:char-IC-IR}}
\begin{proof}
    Let \(\theta, \theta' \in \Theta\).
    Suppose that \(\mathbf{e}\) is implemented by \(d\).
    Then we have 
    \begin{equation}
        \tilde{\pi}(\theta, \mathbf{e}(\theta), \tilde{\mathbf{e}}(d(\mathbf{e}(\theta)))) 
        \geq
        \tilde{\pi}(\theta, e', \tilde{\mathbf{e}}(d(e'))) \label{eq:obedience}
    \end{equation}
    for all \(e'\in E\).
    In particular,
    for \(e' = \mathbf{e}(\theta')\),
    inequality~\eqref{eq:obedience} becomes
    \[
        \tilde{\pi}(\theta, \mathbf{e}(\theta), \tilde{\mathbf{e}}(d(\mathbf{e}(\theta))))
        \geq
        \tilde{\pi}(\theta, \mathbf{e}(\theta'), \tilde{\mathbf{e}}(d(\mathbf{e}(\theta')))).
    \]
    Since \(\mathbf{e} = \tilde{\mathbf{e}}\circ d\circ \mathbf{e}\),
    we have
    \begin{align*}
        \pi(\theta, \mathbf{e}(\theta))
        &= 
        \tilde{\pi}(\theta, \mathbf{e}(\theta), \mathbf{e}(\theta)) \\
        &= \tilde{\pi}(\theta, \mathbf{e}(\theta), \tilde{\mathbf{e}}(d(\mathbf{e}(\theta)))) \\
        &\geq \tilde{\pi}(\theta, \mathbf{e}(\theta'), \tilde{\mathbf{e}}(d(\mathbf{e}(\theta')))) \\
        &= \tilde{\pi}(\theta, \mathbf{e}(\theta'), \mathbf{e}(\theta')) \\
        &= \pi(\theta, \mathbf{e}(\theta')),
    \end{align*}
    which yields inequality~\eqref{eq:IC}.

    Moreover, inequality~\eqref{eq:obedience} holds for \(e' = \bar{e}\),
    which gives
    \begin{align*}
        \pi(\theta, \mathbf{e}(\theta))
        &= \tilde{\pi}(\theta, \mathbf{e}(\theta), \mathbf{e}(\theta)) \\
        &= \tilde{\pi}(\theta, \mathbf{e}(\theta), \tilde{\mathbf{e}}(d(\mathbf{e}(\theta)))) \\
        &\geq \tilde{\pi}(\theta, \bar{e}, \tilde{\mathbf{e}}(d(\bar{e}))) \\
        &= \tilde{\pi}(\theta, \bar{e}, \bar{e})\\
        &= \pi(\theta, \bar{e}),
    \end{align*}
    which yields inequality~\eqref{eq:IR}.

    Now suppose that an emission scheme \(\mathbf{e}:\Theta\to E\) satisfies both \eqref{eq:IC} and~\eqref{eq:IR}.
    Define the function
    \(\tilde{\mathbf{e}}\) as the identity mapping from \(E\) to \(E\).
    Let
    \(d(e) = e\) if \(e \in \mathbf{e}(\Theta)\),
    and \(d(e) = \bar{e}\) otherwise.
    It is straightforward to verify that
    \(\tilde{\mathbf{e}}(d(\mathbf{e}(\theta)))=\mathbf{e}(\theta)\) for all \(\theta\in \Theta\),
    and \((\mathbf{e}, \tilde{\mathbf{e}})\) satisfies the belief consistency requirement.
    It remains to show that
    \(\mathbf{e}\) solves the firm's optimization problem.

    Fix \(\theta\in\Theta\)
    and consider any \(e'\in E\).
    If \(e' = \mathbf{e}(\theta')\) for some \(\theta'\in\Theta\),
    then we have
    \begin{align*}
        \tilde{\pi}(\theta, \mathbf{e}(\theta), \tilde{\mathbf{e}}(d(\mathbf{e}(\theta)))) &= \tilde{\pi}(\theta, \mathbf{e}(\theta), \mathbf{e}(\theta))\\
        &\geq \tilde{\pi}(\theta, \mathbf{e}(\theta'), \mathbf{e}(\theta'))\\
        &= \tilde{\pi}(\theta, e', e')\\
        &= \tilde{\pi}(\theta, e', \tilde{\mathbf{e}}(d(e'))),
    \end{align*}
    where the inequality comes from~\eqref{eq:IC}.
    Otherwise, if \(e'\neq \mathbf{e}(\theta')\) for any \(\theta'\in\Theta\),
    then we have
    \begin{align*}
        \tilde{\pi}(\theta, \mathbf{e}(\theta), \tilde{\mathbf{e}}(d(\mathbf{e}(\theta))))
        &= \tilde{\pi}(\theta, \mathbf{e}(\theta), \mathbf{e}(\theta))\\
        &\geq \tilde{\pi}(\theta, \bar{e}, \bar{e})\\
        &\geq \tilde{\pi}(\theta, e', \bar{e})\\
        &= \tilde{\pi}(\theta, e', \tilde{\mathbf{e}}(d(e'))),
    \end{align*}
    where the first inequality comes from~\eqref{eq:IR},
    and the second inequality comes from the monotonicity of \(\tilde{\pi}(\theta, \,\cdot\,, \tilde{e})\).
    Hence, \(\mathbf{e}\) is implemented by \(d\).
\end{proof}

\printbibliography

\end{document}